\documentclass[useAMS,usenatbib]{mn2e}

\usepackage[squaren]{SIunits}
\usepackage{graphicx}

\title[3D Galactic dust extinction mapping]{3D Galactic dust extinction mapping with multi-band photometry}
\author[R. J. Hanson and C. A. L. Bailer-Jones]{R. J. Hanson$^{1}$\thanks{E-mail:
hanson@mpia.de} and C. A. L. Bailer-Jones$^{1}$\\
$^{1}$Max-Planck-Institut f\"ur Astronomie, K\"onigstuhl 17, 69117 Heidelberg, Germany}
\begin{document}


\pagerange{\pageref{firstpage}--\pageref{lastpage}} \pubyear{201X}

\maketitle

\label{firstpage}

\begin{abstract}
We present a method to simultaneously infer the interstellar extinction parameters $A_0$ and $R_0$, stellar effective temperature $T_{\rm eff}$, and distance modulus $\mu$ in a Bayesian framework. Using multi-band photometry from SDSS and UKIDSS, we train a forward model to emulate the colour-change due to physical properties of stars and the interstellar medium for temperatures from $4\,000$ to $\unit{9\,000}{K}$ and extinctions from $0$ to $\unit{5}{mag}$. We introduce a Hertzsprung-Russel diagram prior to account for physical constraints on the distribution of stars in the temperature-absolute magnitude plane. This allows us to infer distances probabilistically. Influences of colour information, priors and model parameters are explored. Residual mean absolute errors (MAEs) on a set of objects for extinction and temperature are $\unit{0.2}{mag}$ and $\unit{300}{K}$, respectively, for $R_0$ fixed to $3.1$. For variable $R_0$, we obtain MAEs of $\unit{0.37}{mag}$, $\unit{412.9}{K}$ and $0.74$ for $A_0$, $T_{\rm eff}$ and $R_0$, respectively. Distance moduli are accurate to approximately $\unit{2}{mag}$. Quantifying the precisions of individual parameter estimates with $68\%$ confidence interval of the posterior distribution, we obtain $\unit{0.05}{mag}$, $\unit{66}{K}$, $\unit{2}{mag}$ and $0.07$ for $A_0$, $T_{\rm eff}$, $\mu$ and $R_0$, respectively, although we find that these underestimate the accuracy of the model. We produce two-dimensional maps in extinction and $R_0$ that are compared to previous work. Furthermore we incorporate the inferred distance information to compute fully probabilistic distance profiles for individual lines of sight. The individual stellar AP estimates, combined with inferred 3D information will make possible many Galactic science and modelling applications. Adapting our method to work with other surveys, such as Pan-STARRS and Gaia, will allow us to probe other regions of the Galaxy.
\end{abstract}


\section{Introduction}
From our vantage point within the Galaxy, most astronomical observations are affected by interstellar dust, which attenuates light from distant objects as a variable function of wavelength. This is a central problem for many applications, as we typically want to figure out the intrinsic properties of the source, rather than how its emitted photons are influenced by the intervening matter distribution.

The leftovers from nuclear burning of stars, be they emitted through winds or, more energetically through supernovae explosions, produce dust grains that get mixed and reprocessed in interstellar space to eventually form new stars and planets. Dust typically affects electromagnetic radiation from ultraviolet to far-infrared wavelengths, absorbing these photons and re-emitting them as thermal radiation.

Early dust maps used the correlation between the dust column density and the distribution of neutral hydrogen \citep{1978ApJ...225...40B}. This data was supplanted by the dust maps produced by \citet[hereafter SFD]{1998ApJ...500..525S}, who used full sky microwave data made available by the \textit{IRAS} satellite and the \textit{DIRBE} instrument on the \textit{COBE} mission. Mapping the dust column densities via the calibrated dust temperature, the extinction maps, assuming a standard reddening law, were shown to be at least twice as accurate as those of \citeauthor{1978ApJ...225...40B}.

With the advent of large photometric and spectroscopic datasets and the evolution of computing power and data analysis techniques, extinction can now be computed for millions of stars in a reasonable amount of time. Several different methods have been used to do this to date. One can for example determine the reddening vector for a star by computing the shift in colour-magnitude diagrams with respect to unreddened data according to an adopted extinction law. \cite{2010ApJ...725.1175S} use the blue tip of the stellar locus to measure the colour shift of the main-sequence turnoff, which is a good proxy for extinction. In principle interstellar extinction can also be estimated using the spectral energy distribution (SED) of stellar templates directly \citep[e.g.,][]{2012ApJ...757..166B}, often performed via a $\chi^2$ fit to the data. The strong degeneracy between extinction and effective temperature of a star in optical and near-infrared bands affects such approaches, as they typically do not take into account the covariance of the data, conditioned on the APs. 

\cite{2011A&A...534A...3G, 2012A&A...543A..13G} base their maps on measuring the mean $(J-K_{\rm s})$ colour of red clump giants and compare that to the colour of stars measured in Baade's Window for fixed $E(B-V)$. This way they achieve a high resolution map of the central bulge that is insensitive to differential extinction. \cite{2006A&A...453..635M} use a Galaxy model for intrinsic colours and distances and then base their calculations on the near-infrared colour-excess method (\textit{NICE}, \citealt{1994ApJ...429..694L} and \textit{NICER}, \citealt{2001A&A...377.1023L}). Despite unavoidable model dependencies, their results are quite robust to moderate changes in their assumptions.

In our method we use Bayes' theorem to probabilistically characterise the degeneracy between extinction and effective temperature, taking into account covariances in colour space. We estimate the combined probability distribution (PDF) of extinction, effective temperature, distance modulus and $R_0$ for each star individually from which we determine the three-dimensional spatial distribution of dust in the volume probed by our data. This approach expands on a previous study by \cite{2011MNRAS.411..435B}, who used known parallaxes from \textit{Hipparcos} and effective temperatures from spectroscopy to infer temperature and line-of-sight extinction for a smaller set of nearby stars.

Methodologically, our approach is most similar to \cite{2012MNRAS.427.2119S}, who uses a hierarchical Bayesian model to determine the mean distance-extinction relationship for individual lines of sight as well as estimating further stellar parameters. Our method assumes a less complex model to map extinction and distances, thereby avoiding several potential systematic errors resulting from poor choices of priors.

This paper is organised as follows. Section~\ref{sec:theory} covers the basic theory of our approach and describes the Bayesian method used, introducing the individual factors that make up the posterior probability. We illustrate the practical model fitting procedure in Section~\ref{sec:model}, detailing selection and features of the data. We quantify the performance of the method in Section~\ref{sec:validation}, contrasting overall model accuracy with individual precisions. We apply our method to a large sample of data with unknown parameters in Section~\ref{sec:application} in order to investigate the distribution of dust in the Galaxy. We discuss our results and conclude in Section~\ref{sec:summary}.
\section{Theory}
\label{sec:theory}
Our aim is to determine the full probability density function (PDF) over the astrophysical parameters of a star given a set of photometric colours, $\bmath{p}$. Interstellar extinction, $A_{\rm 0}$ (see definition below), is of particular interest. Predicting this parameter requires simultaneous estimation of the effective temperature due to their degeneracy in respect to colour. By introducing a Hertzsprung-Russel diagram (HRD) prior we can also infer an estimate of the distance modulus, which in a band $X$ is $\mu = m_{X} - M_{X} - A_{X} \, ( = 5\log d - 5)$, where $d$ is the distance in $\rm parsec$, $m_{X}$, $M_{X}$ and $A_{X}$ are apparent magnitude, absolute magnitude and extinction, respectively. Using this, we build a forward model to infer the APs $\{A_0, T_{\rm eff}, M_{r}\}$, i.e. estimating $r$-band absolute magnitude, from which we can then compute the distance. We can also estimate the parameter $R_0$, which is the ratio of total to selective extinction. The derivation of an expression for the posterior PDF $P(\mathbf\Theta | \bmath{p}$), where $\mathbf\Theta$ is the list of APs, is presented in this section. This work builds on and adapts the method introduced in \cite{2011MNRAS.411..435B}.
\subsection{Interstellar extinction}
Following \citet{1988ApJ...329L..33C}, the extinction in a narrow band at wavelength $\lambda$ can be parametrised by the extinction parameter $A_{\rm 0}$ and $R_{\rm 0}$ as
\begin{equation}
A_{\lambda} = A_{\rm 0} \cdot ( a_{\lambda} + b_{\lambda} / R_{\rm 0} )\ ,
\end{equation}
where $a_{\lambda}$ and $b_{\lambda}$ are fixed polynomials. Note that we use $A_{\rm 0}$ to characterise the extinction solely as a property of the interstellar medium rather than as a function of the spectral energy distribution of a star, which is strongly dependent on the intrinsic parameters of a particular star.

We introduce artificial reddening using the extinction curves from \citet{1999PASP..111...63F}. Generally, the parameter $R_{\rm 0}$ can be fixed, often at $R_{\rm 0} = 3.1$, which in reality is only a mean value for the diffuse interstellar medium \citep[e.g.][]{1979ARA&A..17..73S, 1979MNRAS.187P..73S}, rather than any individual line of sight in the Galaxy. As we will see in section \ref{subsec:r} it can be advantageous to keep this parameter variable, and sample over it instead by including it in $\mathbf\Theta$.

Extinction in a band $X$ is then computed by integrating over the spectral energy distribution of a star, $F_{\lambda}\equiv F_{\lambda}(T_{\rm eff})$, with fixed effective temperature
\begin{equation}
A_X = -2.5\log\left(\frac{\int\! F_{\lambda} h_{\lambda,X} 10^{-0.4A_{\lambda}}\,{\rm d}\lambda}{\int\! F_{\lambda} h_{\lambda,X}\,{\rm d}\lambda}\right) \ ,
\end{equation}
\noindent where $h_{\lambda,X}$ is the pass band function of filter $X$. We use this to model (predict) the changes in magnitude due to extinction and effective temperature in any band for which we have the pass band.
\subsection{Posterior}
\label{sec:posterior}
We now derive the posterior PDF. The full posterior $P(A_0, T_{\rm eff}, M_{r}|\, \bmath{p}, m_{\rm r}, H)$ quantifies the probability of the parameters, given the measured colours $\bmath{p}$, the apparent $r$-band magnitude $m_{\rm r}$, and HRD $H$. Using Bayes' theorem we can write it as ($T=T_{\rm eff}$, $M=M_{\rm r}$, $m=m_{\rm r}$)
\begin{equation}
 P(A_0, T, M | \bmath{p}, m, H) = \frac{P(\bmath{p}, m | A_0, T, M , H)\,P(A_0, T, M  | H)}{P(\bmath{p}, m | H)} \ .
\end{equation}

The first term is the likelihood, the second term summarises the priors on extinction, temperature and absolute magnitude. We do not need to quantify $P(\bmath{p}, m)$ (the evidence). As it does not discriminate between APs, it is essentially a normalisation constant in the present context. The other terms are described in the following sections.

Not all of the parameters and measurements are dependent on each other, this allows us to rewrite the posterior as
\begin{eqnarray}
 &&P(A_0, T, M |\, \bmath{p}, m, H) \nonumber \\
 &&= \frac{P(\bmath{p} | A_0, T, M , H)\,P(m | A_0, T, M , H)\,P(A_0, T, M)}{P(\bmath{p}, m | H)} \nonumber \\
 &&= P(\bmath{p} | A_0, T)\, \frac{P(A_0, T, M)}{P(\bmath{p}, m)}\,P(m | A_0, T, M, H) \ .
\end{eqnarray}
All terms bar the final one are straight forward to understand, as we have only removed independent variables. The final term we further decompose in order to introduce explicit dependence on absolute magnitude $M$ and build in the HRD. It formalises the influence of the HRD prior as the marginalisation over $M$:

\begin{eqnarray}
&&P(m | A_0, T, M, H) \nonumber \\
&&= \int\limits_{M}\! P(m | M, T, H)\, P(M | T, H)\,{\rm d}M \nonumber \\
&&= \int\limits_{M}\! P(m | M, T, H)\, \frac{P(M, T | H)}{P(T)}\,{\rm d}M \ ,
\label{eq:hrdpriorint}
\end{eqnarray}
and does not explicitly depend on extinction. $A_0$ drops out of the first term because $m$ is independent of it once conditioned on $M$. The full posterior function is now complete and can be written as
\begin{eqnarray}
&&P(A_0, T, M | \bmath{p}, m, H) = \nonumber \\
&&P(\bmath{p} | A_0, T)\frac{P(A_0)}{P(\bmath{p}, m)}\! \int\limits_{M}\! P(m | M) P(M, T | H) \,{\rm d}M .
\label{eqn:post1}
\end{eqnarray}

Individually, we only impose a uniform prior on absolute magnitude within the parameter boundaries of the HRD. This is implicitly included in the integral already, hence we take it out of the equation here. The other APs have regular uniform priors. 

We can also build another forward model which also takes $R_{\rm 0}$ into account. The corresponding posterior function takes the analogous form

\begin{eqnarray}
P(A_0, T, R_0, M | \bmath{p}, m, H) = \nonumber \\
 P(\bmath{p} | A_0, T, R_0)\frac{P(A_0, R_0)}{P(\bmath{p}, m)} \ \int\limits_{M}\! P(m | M) P(M, T | H) \,{\rm d}M \ ,
\label{eqn:post2}
\end{eqnarray}
where the parameter formally enters in the likelihood and possible priors. In our case, we impose a uniform prior on $R_{\rm 0}$.
\subsection{Forward model}
The forward model predicts the star's colours given a set of astrophysical parameters. It is calculated by fitting a thin plate spline (a multidimensional analogue of a one-dimensional spline) as a function of $A_{\rm 0}$ and $T_{\rm eff}$ to each colour in $\bmath{p}$, $\bmath{f}(A_{\rm 0}, T)=\bmath{p}'$. Although we infer absolute magnitudes with our method, we do not include it in the forward model.

We can furthermore train a forward model which includes $R_0$, expanding the dimensionality of the spline and characterising the variation over all three parameters, altering our function to $\bmath{f}'(A_{\rm 0}, T, R_{\rm 0})$. The posterior in the previous section remains largely unchanged.
\subsection{Likelihood}
\label{subsec:likelihood}
The likelihood of the colours $\bmath{p}$, given the forward model predictions of the colours, $\bmath{f}$ (or $\bmath{f}'$), and assuming Gaussian errors on the measurements, is
\[
P ( \bmath{p} | A_0, T) = k \cdot \exp\left(-0.5\,(\bmath{p}-\bmath{f})^{T} \bmath{C}^{-1}(\bmath{p}-\bmath{f})\right) \ ,
\]
where $k = (2\pi)^{-n/2} |\bmath{C}|^{-1/2}$ is the normalisation factor, dependent on the dimensionality $n$ of the colour vector $\bmath{p}$. The covariance matrix $\bmath{C}$ is determined from the errors on the measurements (magnitudes), taking into account the fact that the errors on consecutive colours are not independent. Taking just the five SDSS bands for illustration, the covariance matrix for the resulting colours $\bmath{p} = (u-g, g-r, r-i, i-z)$ has the following form

\begin{equation}
\bmath{C} = \left( \begin{array}{cccc}
\sigma_u^2 + \sigma_g^2 & -\sigma_g^2 & 0 & 0 \\
-\sigma_g^2 &  \sigma_g^2 + \sigma_r^2 & -\sigma_r^2 & 0 \\
0 & -\sigma_r^2 & \sigma_r^2 + \sigma_i^2 & -\sigma_i^2 \\
0 & 0 & -\sigma_i^2 & \sigma_i^2 + \sigma_z^2 \end{array} \right) \, ,
\end{equation}
where $\sigma_X^2$ is the variance in band $X$. For additional colours, the matrix expands analogously. A standard $\chi^2$ approach would ignore the off-diagonals, treating the colours as independent, which is obviously not correct.

\subsection{HRD Prior}
\label{subsec:priors}
The prior $P(M, T | H)$ is constructed to quantify the relative probability of finding a star in a particular region of the HRD. Due to the nature of stellar evolution, the HRD is inhomogeneously populated. Fixing one parameter constrains the other. This allows us to extract valuable information, even for a crude representation of the HRD (see section \ref{subsec:buildfm} for details on how it is constructed).

To achieve this, we need to construct an equation that links the apparent and absolute magnitudes and distance. By independently estimating extinction and effective temperature from the colours und using the measured $m_{\rm r}$ and the HRD, we can in principle determine the absolute magnitude and from that estimate the distance modulus. To effectively constrain the absolute magnitude from noisy measurements and the finite width of the HRD, we introduce a noise model to determine the probability distribution. We define the random variable
\begin{equation}
\kappa = m_{\rm r} - M_{\rm r} - \Delta \ ,
\end{equation}
where we define $\Delta$ as the difference between the true (but unknown) apparent and estimated absolute magnitudes. For error-free measurements, $\kappa$ is zero. By connecting apparent and absolute magnitudes in this way, $\Delta$ can be considered a proxy for the distance modulus, which we are trying to infer.

For the noise model $P(\kappa | M_{\rm r})$ we select a one-dimensional Gaussian in $\kappa$, $\mathcal{N}(0, \sigma_{m_{\rm r}})$ which has an expectation value of zero and a standard deviation $\sigma_{\Delta} = \sigma_{m_{\rm r}}$, the uncertainty in the $r$-band magnitude measurement for the observed star. The absolute magnitude is not measured and as such has no error, $\Delta$ is sampled. For a particular measurement of $m_{\rm r}$ the probability is
\begin{equation}
P(\kappa|M_{\rm r}) = \mathcal{N}(m_{\rm r}-M_{\rm r}-\Delta,\,\sigma_{m_{\rm r}}) \ .
\end{equation}

As $m_{\rm r}$ is the only noisy quantity in $\kappa$, this term can also be written as $P(m_{\rm r} | M_{\rm r})$. This represents the first term of the integral in equations \ref{eqn:post1} and \ref{eqn:post2}. So, for any measure of $\Delta$, obeying $\kappa = 0$, the Gaussian describes the scatter around this value. The full HRD prior probability is then calculated by integrating the Gaussian for a $\Delta$ over the probability distribution of the HRD at a given $T_{\rm eff}$, as formalised in Equation~\ref{eq:hrdpriorint}.

Using the HRD gives us a self-consistent means to estimate distances from photometric information. A stronger constraint on APs can be constructed if distance information is available. The original work by \cite{2011MNRAS.411..435B} uses Hipparcos parallaxes to constrain the APs in this way.
\section{Model fitting}
\label{sec:model}
In this section we outline the practical implementation of the method and describe the data we use to build and validate the forward model and priors.
\subsection{Data}
To perform the analysis described in Section~\ref{sec:theory}, a set of photometric data is required, for which astrophysical parameters have been independently determined. We use stars selected from the \textit{UKIRT Infrared Deep Sky Survey}, UKIDSS \citep{2007MNRAS.379.1599L} DR9 Plus and \textit{Sloan Digital Sky Survey}, SDSS \citep{2011ApJS..193...29A} DR8. We use data from the UKIDSS \textit{Large Area Survey} (LAS) which is designed to overlap with the footprint of SDSS. By using real data, we are able to use the proper photometric errors and intrinsic scatter to build the forward model, instead of depending on synthetic estimates. We perform a crossmatch to generate a catalogue of stars with photometry in nine bands. The bands \textit{ugriz} are from SDSS and \textit{YJHK} are covered by UKIDSS LAS. In principle we can also use (or combine) other surveys, as our method can easily be trained on further photometric data.
\subsection{Training data for fitting forward model}
\label{subsec:training}
We need to simulate the effects of extinction to build an extended catalogue of stars with known APs. We therefore need to construct a catalogue of (ideally) zero extinction stars. To build our training data we select only stars at high Galactic latitudes, $b > 70^\circ$, which avoids the Galactic plane and most of the interstellar dust. Additionally these stars are required to be dwarfs (surface gravity $\log g > 4$). This ensures that the catalogue stars only have minimal variations in $\log g$, which improves the temperature estimates due to the tight correlation with colour for fixed extinction (as in Figure~\ref{fig:teffcolfixa0}). Furthermore, effective temperatures determined with the \textit{SEGUE Stellar Parameter Pipeline}, SSPP \citep{2008AJ....136.2022L, 2009AJ....137.4377Y} must be available, with errors on these of less than $\unit{200}{K}$.

We use a weighted average of several spectroscopic and photometric temperature estimates (called \texttt{ADOP} in the SSPP papers) for this work. When using instead the purely spectroscopic estimate, e.g. the \texttt{ANNRR} routine  \citep{2007A&A...467.1373R}, we see no significant differences in our inferred extinction and distance estimates. The temperature estimates do differ, with a standard deviation of about $\unit{80}{K}$. But as we are not primarily interested in inferring temperatures, we can in principle use either of the estimates, though using the spectroscopic estimator avoids circular arguments.

We chose the $r$-band magnitudes to be $r < 19$, which typically results in photometric errors approaching the systematic limit of approximately $\unit{0.02}{mag}$. This also generally forces the photometry in the other bands to be complete and of good quality. To further ensure selection of sources with high quality data, we impose the following selection criteria on the SDSS flags: we select only unresolved sources (\texttt{sdss.type} = $6$) with clean photometry (\texttt{sdss.clean} = $1$) and general quality flags set (\texttt{sdss.'4295229440'} = $0$, \texttt{sdss.mode} = $1$). Similarly, we select objects with clean photometry from UKIDSS (\texttt{ukidss.*ppErrBits} $<$ $65\,536$, one for each band) that are classified as stellar objects (\texttt{ukidss.mergedClass} = $-1$).

The data (SDSS and UKIDSS) are queried and cross-matched using the \textit{WFCAM Science Archive}, requiring complete photometry in all bands. In total, we obtain roughly $3\,200$ stars that fulfil all the above criteria. Compared to the hundreds of thousands of stars analysed in SEGUE, we retrieve comparatively few. Aside from completeness and crossmatching reasons, this is mainly influenced by the magnitude, surface gravity and temperature limits we set. In spite of this, this does not introduce systematic biases. Our selection is representative of the full SEGUE dataset (see below). The photometry is then corrected for the small estimated extinction from SFD, where the assumption holds that at high Galactic latitudes the stars are behind all the layers of dust.

\begin{figure}
\includegraphics[width=\columnwidth]{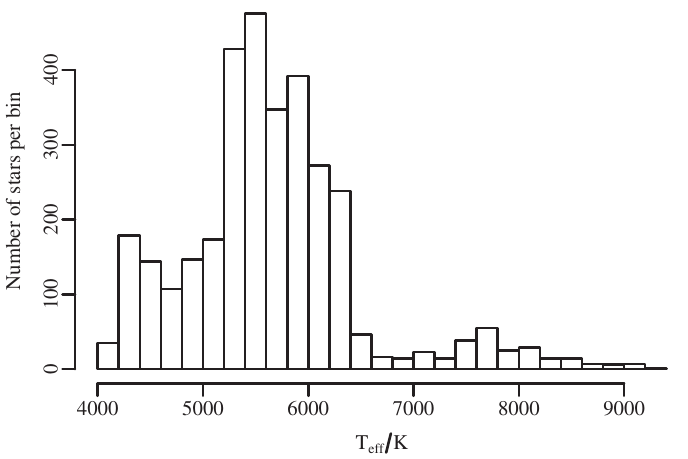}
\caption{Distribution of effective temperatures of approximately $3\,200$ crossmatched stars. Temperatures are estimated by SSPP with mean uncertainties of $\unit{58.5}{K}$. Of these stars, one thousand are randomly chosen to build the extended dataset.}
\label{fig:teffhist}
\end{figure}

Of the selected data, a randomly sampled set of one thousand stars is chosen, covering a temperature range of approximately $\unit{4000 - 9000}{K}$. A histogram of the temperature distribution for the full training set is shown in Figure~\ref{fig:teffhist}, showing a peak around $\unit{5600}{K}$. Only $\unit{9}{\%}$ of the stars have effective temperatures above $\unit{6500}{K}$. This is representative for the whole SEGUE dataset.

When later applying the method to data with unknown stellar parameters, we still create a selection with the quality flags used above, though we naturally impose no requirement on the existence of temperature estimates or the position in the Galaxy (low extinction).
\subsection{Adding artificial extinction}
\label{subsec:artext}
Using the filter response functions for all filters and an extinction law \citep{1999PASP..111...63F}, the change in magnitude for a given extinction $A_0$ in each band is computed for a range synthetic SEDs. Owing to the fact that we are using broad band filters, the extinction in a band varies smoothly with temperature and we do not need to do this for every unique temperature present in our data sample. Instead we use eleven \textit{PHOENIX} model SEDs \citep{1999ApJ...512..377H} evenly spaced in the temperature range from $\unit{4000}{K}$ to $\unit{9000}{K}$. This covers the temperature range in our catalogue. For each band we fit a set of one-dimensional quadratic functions in $T_{\rm eff}$ for fixed $A_{\rm 0}$, covering the whole temperature range noted above and extinction in steps of $\unit{0.25}{mag}$ from $0$ to $5$ magnitudes.

Knowing how the magnitudes vary with $A_0$ and $T_{\rm eff}$, we expand our initial dataset by adding artificial extinction in $21$ discrete steps over the range noted above for each of the $1\,000$ stars. The photometric errors are adapted to take into account the change in magnitudes by following the average magnitude-error relation in the data. Our expanded dataset (\textit{training set}) now comprises $21\,000$ stars with known artificial extinction and effective temperatures.
\subsection{Building the forward model}
\label{subsec:buildfm}
\begin{figure}
\includegraphics[width=\columnwidth]{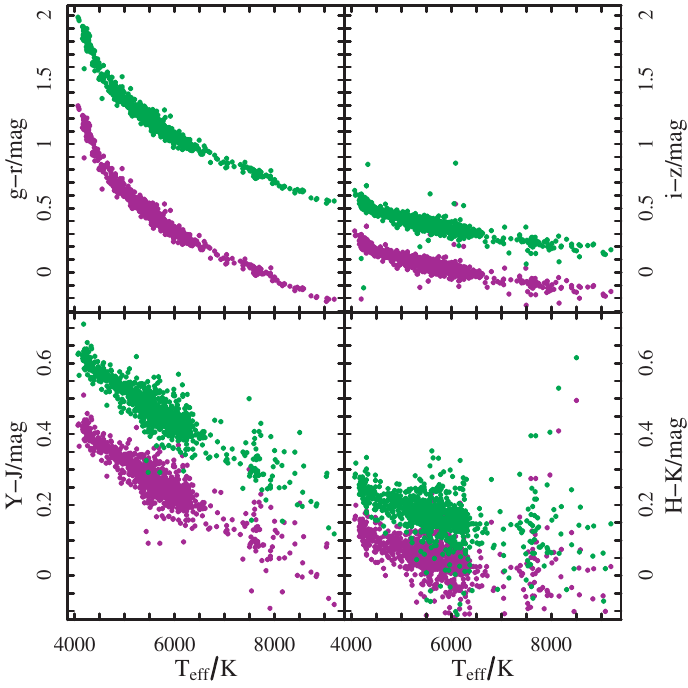}
\caption{Effects of effective temperature and extinction on four selected colours. Purple points show a subset of the training data with no extinction as a function of effective temperature. Green dots show the effects for $A_0 = \unit{2}{mag}$. Each sample consists of $1\,000$ stars.}
\label{fig:teffcolfixa0}
\end{figure}

In Figure~\ref{fig:teffcolfixa0} we show the effect of temperature and extinction on four different colours. We do not convert the magnitudes in both surveys to a common system (UKIDSS uses Vega magnitudes, for example), because the colours are merely affected by a constant offset with respect to that of the other survey, and we are only interested in changes in colours. As long as the training data and the final datasets are handled consistently, we need not take into account this offset.

From the training set we select a random subset of $4\,000$ stars to create the forward model itself, with the aim of being able to predict colours given temperature and extinction (see section \ref{subsec:likelihood}). For each of the eight reduced colours we fit a two-dimensional thin-plate spline to both APs. A good fit for all colours is achieved when giving the splines $20$ degrees of freedom.

\begin{figure}
\includegraphics[width=\columnwidth]{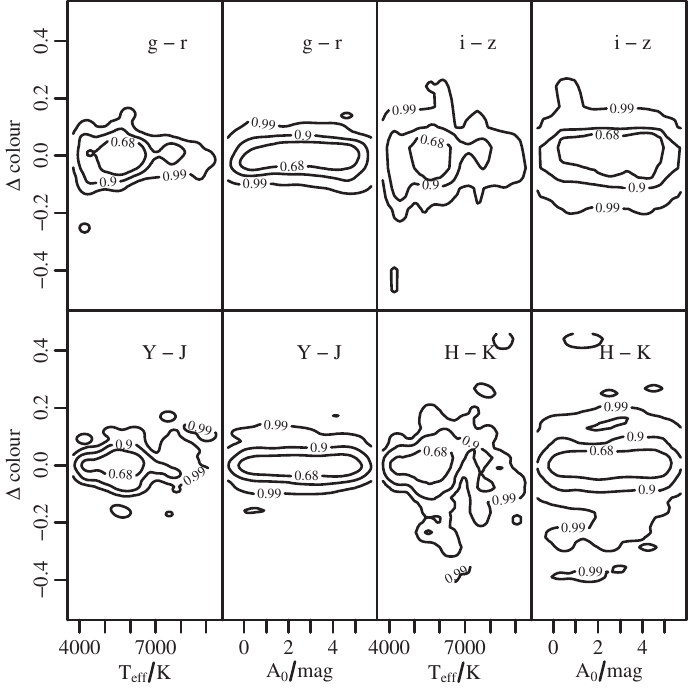}
\caption{Residuals in the forward model. This shows the difference between the true colours and the colours predicted from the forward model thin plate spline as functions of true effective temperature and extinction, for four colours. The contour levels shown incorporate $68\%$, $90\%$ and $99\%$ of the data points. In the top right of each panel the corresponding colour is named, with the x-axis depicting the respective parameter ranges. All panels are scaled to show the same range on the y-axis. Typical scatter is in the order of $\unit{0.03-0.06}{mag}$ for all colours.}
\label{fig:fmtpsresiduals}
\end{figure}

In Figure~\ref{fig:fmtpsresiduals} we show the direct performance of the splines predicting colours given some input APs. For four example colours we plot the residuals of colour ($p_{X, true} - p_{X, spline}$) for the ranges of temperature and extinction present in the model as contours. This is not the final result of the full parameter inference, instead it illustrates the intrinsic inaccuracies of the model which derive from the scatter in the real data. We do not see any systematic variation, and measure a typical scatter in the order of $\unit{0.03-0.06}{mag}$ for all colours, which is roughly $\sqrt{2}$ larger than for the magnitudes in a band. The forward model captures the intrinsic variance, as this is the same order of magnitude or within a small factor of the reported uncertainties of the survey data magnitudes. In Section~\ref{subsec:accapest} we look in detail at how assuming larger photometric errors affects the AP estimation.

\begin{figure}
\includegraphics[width=\columnwidth]{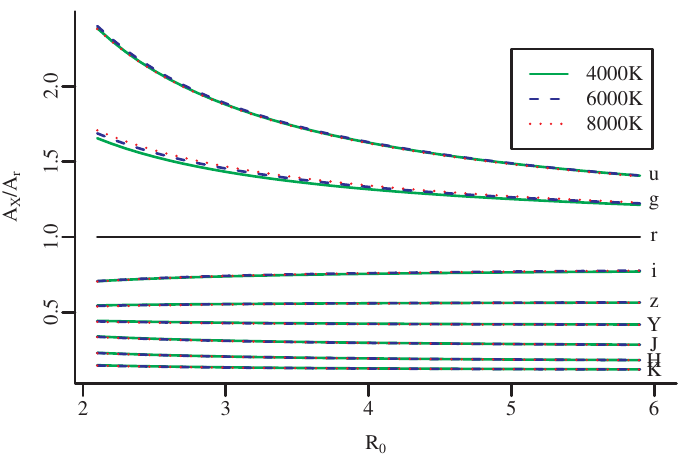}
\caption{Model values for extinction relative to $r$-band for three temperatures over a range of $R_0$ and the photometric bands used in this work. Small temperature dependences can be seen for the optical bands $u$ and $g$ at low $R_0$.}
\label{fig:extr0teffva}
\end{figure}

We characterise extinction using $A_0$ to indicate that this is a parameter that solely depends on the physics of the interstellar medium, rather than on the type of star. When presenting results and maps later on we will use extinction in SDSS $r$-band to allow more convenient comparisons to other work. Note that the conversion from $A_0$ to $A_{\rm r}$ has a small dependence on temperature. For $R_0 = 3.1$ they roughly convert as $\left<A_{\rm r}/A_0\right> = 0.837$. The standard deviation from this value over the parameter range $T_{\rm eff} = \unit{4000-9000}{K}$ and $A_0 = \unit{0-5}{mag}$ is only $0.005$. The model values for extinction in other bands, e.g. the relative extinctions $A_X/A_{\rm r}$ are listed in Table~\ref{tab:modelextval} for three different values of $R_0$. Values reported are averages over the above range. In particular, changes occur for the short wavelength optical bands. When varying $R_0$, we see that the relative extinction in the other, redder, bands changes only slightly. In Figure~\ref{fig:extr0teffva} these changes are illustrated for all the bands used and for three different effective temperatures.

\begin{table}
\caption{Model values for extinction in all bands, compared to SDSS $r$-band extinction, $A_X/A_{\rm r}$ for three values of $R_0$. Reported values are averages over the full temperature range of the data.}
\label{tab:modelextval}
\centering
	\begin{tabular}{ccccccccc}
	$R_0$	& $u$	& $g$	& $i$	& $z$	\\
	2.1		& 2.393	& 1.698	& 0.707	& 0.542	\\
   	3.1		& 1.848	& 1.445	& 0.744	& 0.555	\\
	4.1		& 1.610	& 1.329	& 0.761	& 0.561 	\\ \hline
	$R_0$		& $Y$	& $J$	& $H$	& $K$ \\
	2.1		& 0.439	& 0.337	& 0.230	& 0.147 \\
	3.1		& 0.427	& 0.308	& 0.204	& 0.133  \\
	4.1		& 0.422	& 0.296	& 0.193	& 0.126 \\
  \end{tabular}
\end{table}

For the APs used to build the forward model we implicitly implement priors that are uniform over the APs' training ranges, and have zero probability outside them. We will demonstrate the use of a prior on extinction based on SFD estimates later in Section~\ref{subsec:furtherpriors}.
\subsection{HRD prior}
\label{subsec:hrdprior}
The HRD prior is constructed using stellar isochrones from \citet{2010Ap&SS.328..331S}. The chosen stellar population comprises $100\,000$ stars drawn from a Salpeter initial mass function with masses ranging from $\unit{0.2-107}{M_{\sun}}$ (though we only use stars with $T_{\rm eff} \leq \unit{9\,000}{K}$ and the vast majority of stars have masses below $\unit{1.6}{M_{\sun}}$). Assuming a constant star formation rate over the age of the Universe ($\unit{13.7}{Gyr}$), all stars evolve independently with solar metallicity, i.e. there is no chemical enrichment. By using the forward model, however, we are implicitly assuming that the metallicity of stars we encounter have the same mean metallicity of the SSPP stars, which are used to fit the model. This mismatch between the metallicities of the forward model and the HRD does not strongly affect the estimation of $A_0$ and $T_{\rm eff}$ directly, though it does limit the ability of getting good distance estimates. As we are not estimating metallicity, this mismatch will naturally occur for most stars, independently of the choice of HRD metallicity. An example of this limitation is given in Section~\ref{subsec:demo}.

\begin{figure}
\includegraphics[width=\columnwidth]{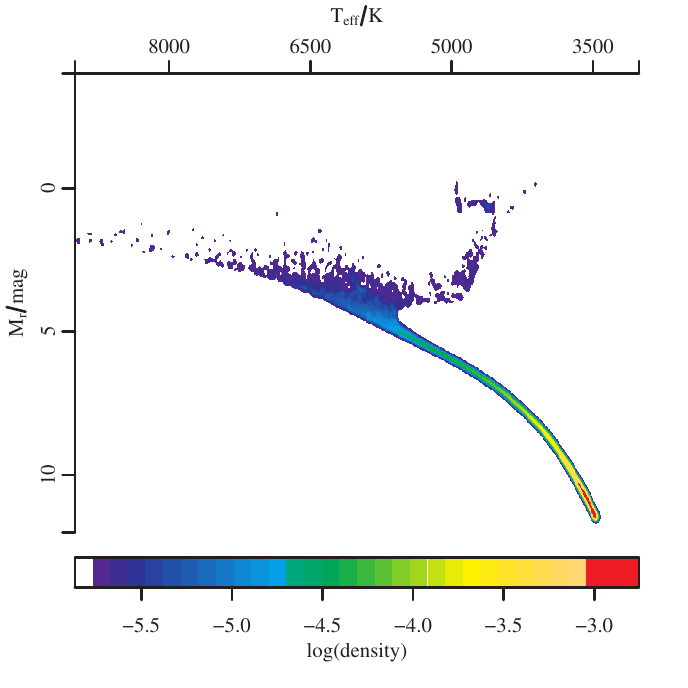}
\caption{Density representation of the HRD used in this analysis. The integrated probability of the HRD is normalised to one, red is high number density and blue is low. The colour scale shows base $10$ logarithm of the density. White areas denote regions of the parameter space with initially zero probability. A small offset is added to each point before normalisation to avoid this in the actual computation. In this case the offset is approximately $10^{-3}$ times the maximum density.}
\label{fig:hrd}
\end{figure}

The resulting temperatures and absolute magnitudes then are used to place the stars in the HRD plane. To use the HRD as a prior, we need to convert the data representation to a probabilistic description of the distribution. We achieve this by applying a two-dimensional binned kernel density estimate with a bandwidth of $\unit{12.5}{K}$ and $\unit{0.0625}{mag}$ in effective temperature and $r$-band absolute magnitude, respectively. The resulting grid has the pixel dimensions of $600^2$, we limit the temperature range to $\unit{3010 - 9000}{K}$ as to extend slightly beyond the limits of the forward model. This allows for the main sequence to be fully represented down to low temperatures without an artificial cutoff. The resulting magnitude range is $r \in \unit{(-4,12)}{mag}$. Before normalisation, a small, but non-zero, offset is added to each point to account for the regions which nominally have zero probability, but in reality may not be completely empty. These regions include white dwarfs in the lower left part of the figure or (post) AGB stars higher up the branch to the right. Statistically, these regions will be very thinly populated and we don't expect to find many (if any) of these stars in our samples, so we don't model these in a more sophisticated manner. A representation of the final distribution is presented in Figure~\ref{fig:hrd}.

\subsection{Sampling and computation}
\label{subsec:sampling}
The forward model, priors and data are combined to compute the posterior function, as detailed in Section~\ref{sec:posterior}. As we are interested in inferring multiple APs, we use a standard Metropolis-Hastings Markov-Chain Monte-Carlo (MCMC) routine to efficiently sample the parameter space and to compute the posterior probability distribution. The parameter space is explored in logarithmic units of the respective quantity, forcing the estimates to remain positive and physical. The sampling matrix is diagonal with variances of $\unit{0.1}{dex}$ in each variable (AP). We assume no correlation between the variables. Convergence is sufficient with a chain length and burn in of  $10\,000$ steps each. This does not need to be adjusted if we also infer $R_0$.

To compute the likelihoods we can either use the actual spline function or build a lookup table. Considering the numerical speed-up we achieve by tabulating results in advance, we generally use this approach. For this, we build a grid in extinction and effective temperature (and later in $R_{\rm 0}$ too) and calculate the predicted colours. We use the same parameter ranges as for training (avoiding extrapolation into regions not covered by the data), creating a grid with $A_0$ from $\unit{0-5}{mag}$ in steps of $\unit{0.001}{mag}$ and $T_{\rm eff}$ from $\unit{3000 - 9000}{K}$ in steps of $\unit{10}{K}$ (effectively the same as for the HRD). For $R_0$ we use a coarser stepsize of $\unit{0.2}{}$ in the range from $2.1-5.9$. The stepsizes are chosen as to be significantly smaller than the mean absolute error (MAE) of the residuals when using the full spline function (see Section~\ref{subsec:effectcolours}.)

When assessing the performance in the APs and computing further quantities, we only use those stars whose inferred parameters are within the boundaries of the lookup table (i.e. within our training space). As indicated in section \ref{subsec:training}, at lower temperatures and smaller wavelengths the linear relation between colour and temperature for fixed extinction no longer holds, therefore extrapolation is not advisable. This avoids skewing the results due to boundary effects, like MCMC chains sticking to the edge of the grid in one or more parameters.

We apply this filter after parameter inference as we cannot rule out including stars that have true APs outside of our training range (we do not apply any colour-cuts to the data selection). If we were to instead force a prior over the same ranges during sampling we would artificially force those stars to have incorrect parameters.

Typically this filter throws out roughly $\unit{5}{\%}$ of stars when estimating extinction, effective temperature and distance modulus. When including the $R_0$ parameter, this increases to close to $\unit{30}{\%}$, due to the additional variation in the model.
\section{Validation}
\label{sec:validation}
In this section we first demonstrate the viability of the model and then analyse the precision and accuracy of the method, as well as the colours. The following analysis is performed on a different set of $4\,000$ stars (\textit{validation set}), of which none where used to train the model.
\subsection{Simulated photometry}
\label{subsec:demo}
\begin{figure}
\includegraphics[width=\columnwidth]{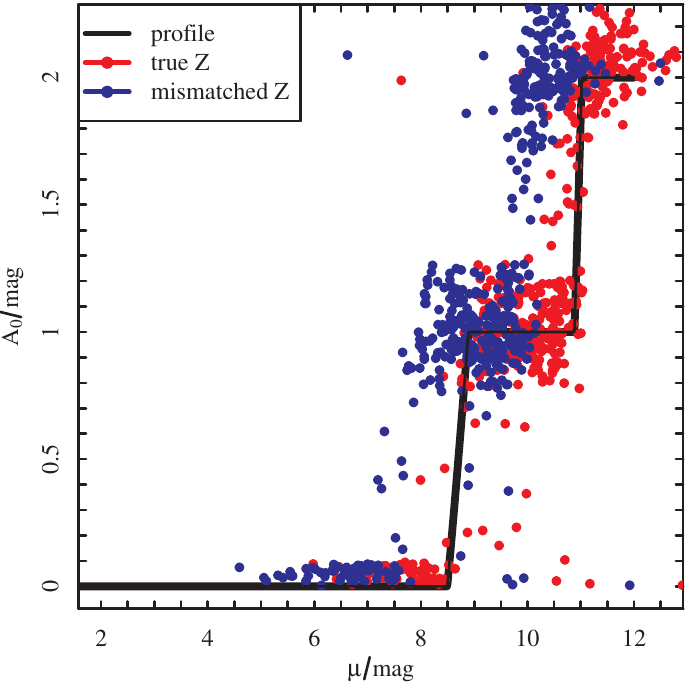}
\caption{Inferred distance moduli $\mu$ and extinction $A_0$ for $500$ stars. Red dots indicate results for which matching HRD metallicity ($Z=0.019$) is used for sampling and inference. Blue dots show results for mismatched HRD ($Z' = 0.0019$).}
\label{fig:demo}
\end{figure}

In this first step we aim to show the expected performance of the method and illustrate its limitations with respect to metallicity (see Figure~\ref{fig:demo}). For this purpose we set up two extinction clouds at $\unit{500}{pc}$ ($\mu\approx\unit{8.5}{mag}$) and $\unit{1500}{pc}$ ($\mu\approx\unit{10.9}{mag}$) with extensions of $\unit{100}{pc}$. This true extinction distribution with $\mu$ is shown as the solid black line. We then place $500$ stars along the line of sight, generating effective temperatures and absolute magnitudes from the HRD prior (solar metallicity). According to their effective temperatures, extinctions and distances, we simulate appropriate photometry and photometric errors, which are characteristic of the SDSS/UKIDSS sample we use. Finally we use our method to infer these parameters. These are plotted as points in the figure.

We infer the APs for two cases. In (a) we use the same solar metallicity HRD prior ($Z=0.019$) as for selecting the stars (red dots). In this case, we can see that our method recovers the distances and extinctions very well, with points closely tracing the extinction profile. The few points that lie furthest from the true profile typically have bad temperature estimates, resulting in erroneous distance moduli. In (b) we use an HRD with a tenth of that metallicity ($Z'=0.0019$, blue dots) to infer parameters. Although the individual estimates for extinction and temperature are compatible, the mismatched stars are systematically shifted by $\Delta\mu = -1$ to closer distances. Analogously, the opposite shift occurs when reversing the mismatch. Therefore the accuracy of distance estimates is strongly limited by the unknown metallicity of each individual star.

It is difficult to estimate metallicity using only broad band photometry, as used here, when neither temperature nor extinction are fixed. As shown in \cite{2013arXiv1309.2157B} for simulated Gaia data, given spectroscopic information it is possible to estimate metallicity simultaneously, resulting in more accurate distance estimates.

\subsection{Effects of Colours and NIR data}
\label{subsec:effectcolours}
Considering that we are using photometry from two surveys covering optical and near-infrared bands, it is instructive to compare how inclusion (or exclusion) of certain bands affects the parameter estimation. To examine this we train the forward model using (a) all nine bands, resulting in the eight colours \textit{u-g}, \textit{g-r}, \textit{r-i}, \textit{i-z}, \textit{z-Y}, \textit{Y-J}, \textit{J-H} and \textit{H-K}, (b) using only the five SDSS bands and thus only four colours and (c) using all but the SDSS $u$ band magnitudes. Generally, one would expect the use of more colours (more information) to yield better results, particularly when taking into account that NIR wavelengths are less affected by dust attenuation than shorter ones.

As expected, the performance of the method is significantly better when using the NIR data (using only NIR data is also much worse). Table~\ref{tab:residuals} summarises the differences between the two cases. We characterise the performance with three key statistics. The first is the bias, which is purely the mean value of the difference between predicted and true parameter for all stars. The second is the mean absolute error (MAE), which similarly is the mean of the absolute difference and the third is the root mean square (RMS). For temperatures below $\unit{6500}{K}$ the performance is generally better than for higher temperatures. This results from the combination of having fewer stars in the higher temperature range and the scatter in the data being larger there (see Figures~\ref{fig:teffhist} and \ref{fig:teffcolfixa0}).

\begin{table}
\caption{Performance using all colours (a), just the four SDSS colours (b) and all bands except $u$ (c). Table shows performance characterised by bias, mean absolute error (MAE) and root mean square (RMS) of the residuals, i.e. the difference between estimated and true parameters. $R_0=3.1$ is fixed in these models.}
\label{tab:residuals}
\centering
	\begin{tabular}{ccccccc}
				&	\multicolumn{3}{c}{$A_0$/mag}		&	\multicolumn{3}{c}{$T_{\rm eff}$/K}	\\
				&	(a)			&	(b)		&	(c)		&	(a)		&	(b) 		&	(c)						\\ \hline
   		bias		& 	$-0.08$		&	$-0.41$ 	& 	$-0.29$	&	$-67$	&	$-295$ 	&	$-269$	\\
		MAE		& 	$0.23$		&	$0.59$ 	&	$0.41$	&	$300$	&	$628$ 	&	$440$	\\
		RMS		& 	$0.45$		&	$1.05$ 	&	$0.68$	&	$586$	&	$1028$ 	&	$712$	\\
  \end{tabular}
\end{table}

\begin{figure}
\includegraphics[width=\columnwidth]{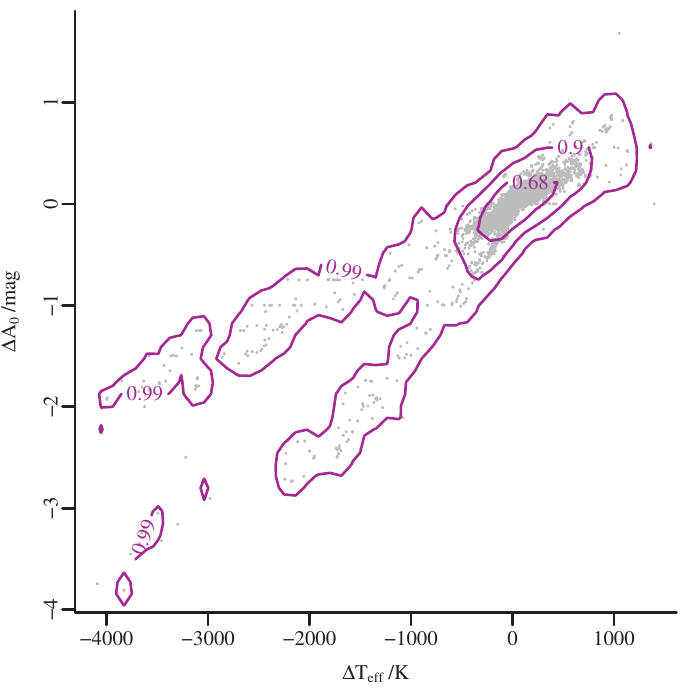}
\caption{Residuals in effective temperature and extinction for case (a) using all eight colours. Plotted are the $68\%$, $90\%$ and $99\%$ density contours of the distribution of points in the plot. The differences are computed as $\Delta\Theta = \Theta_{\rm estimated} - \Theta_{\rm true}$ for $A_0$ and $T_{\rm eff}$.}
\label{fig:residuals}
\end{figure}

We note that the bias is negative in both cases and parameters, though much closer to zero in case (a). Intrinsic scatter in the data also precludes higher residual accuracy. In Figure~\ref{fig:residuals} we plot the results for the validation set of stars. Towards negative values in both parameters we see two distinct tails. Stars in the tails tend to have hotter true temperatures and larger intrinsic scatter. Only a few percent of stars are effected by such a large error in the parameter estimation. Without these tails, the performance would improve only minimally in MAE and RMS.

When excluding only the $u$ band, case (c), we find that the performance is still better than case (b) although significantly worse than (a). The bias, MAE and RMS for $A_0$ are $\unit{-0.29}{mag}$, $\unit{0.41}{mag}$ and $\unit{0.68}{mag}$, respectively. For $T_{\rm eff}$ these are $\unit{-269}{K}$, $\unit{440}{K}$ and $\unit{712}{K}$.

In temperature and extinction we see no significant systematic effect due to the metallicity of the HRD. As indicated in the previous section, this mainly affects the distance estimator.

\begin{figure}
\includegraphics[width=\columnwidth]{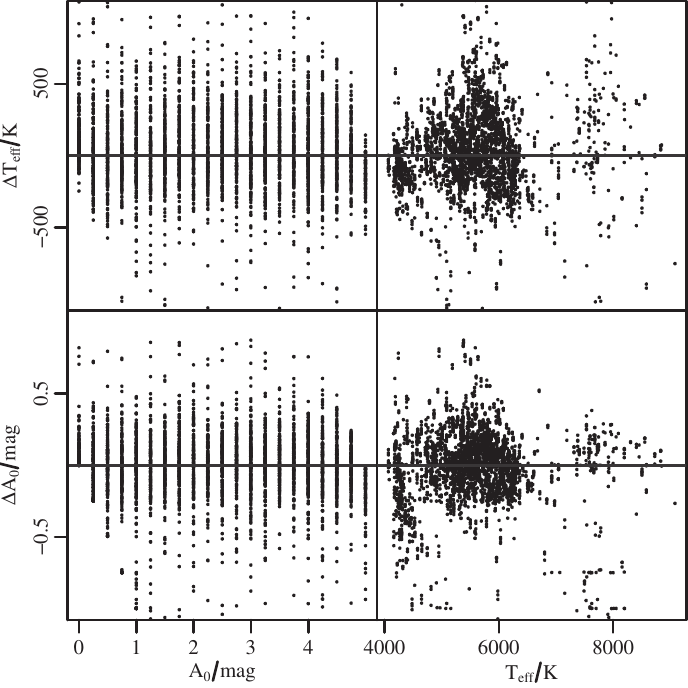}
\caption{Residuals in effective temperature and extinction as functions of the true values in both parameters, as applied to the validation set of data. All eight colours are used. The true variation is shown for the same data as in Figure~\ref{fig:residuals}. A few dozen points are cut off by the selection of the axis range.}
\label{fig:residuals2}
\end{figure}

Figure~\ref{fig:residuals2} shows the residuals for both parameters as a function of the true parameters. We see no systematic behaviour, though we see a general bias towards negative values, which is evident in the values reported in Table~\ref{tab:residuals}.

As we do not have any true distances to compare the inferred distances to, we can not reference any results for distance modulus here. We must rely on the other parameters to be accurate in order to believe the distance results we obtain. We can however roughly estimate distances using the true extinctions and temperatures and finding the corresponding $\Delta = m_{\rm r} - M_{\rm r}$ via the HRD. Doing this we compute a bias of $\unit{-0.23}{mag}$, MAE of $\unit{2.29}{mag}$ and RMS of $\unit{3.31}{mag}$ when using all bands. Approximately $\unit{15}{\%}$ of the estimates are better than  $\unit{0.2}{mag}$. In this case, the distance is poorly constrained by the combination of photometric data and our broad HRD prior, when $T_{\rm eff}$ and $A_0$ also have to be estimated from the data. The bias we encounter is primarily due to the mismatch of metallicities of the HRD and training data.

\subsection{Precision of AP estimation}
\label{subsec:accapest}
\begin{figure}
\includegraphics[width=\columnwidth]{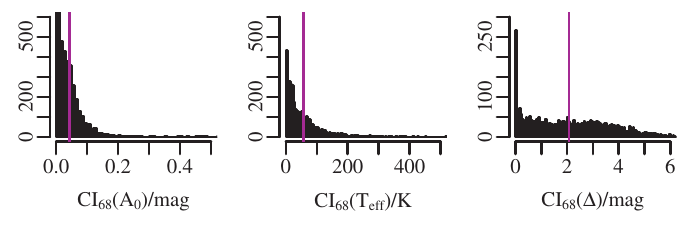}
\caption{Histograms of $68\%$ confidence intervals for extinction and temperature using photometric errors in the covariance matrix. The purple vertical lines indicate the mean values in each case. For $A_0$ this corresponds to $\unit{0.04}{mag}$, for $T_{\rm eff}$ it is $\unit{58}{K}$ and for $\Delta$ it is $\unit{2.0}{mag}$.}
\label{fig:accuracy}
\end{figure}

Beyond AP estimation accuracy (of the method), we are also interested in the precision, i.e. what are our parameter uncertainties for any given star. We quantify this using the width of the $68\%$ confidence interval ($\text{CI}_{68}$) of the posterior PDF (which is equivalent to the $1\sigma$ variation), using a Gaussian kernel density estimation with the mean as the central point. The distribution is not generally symmetric, so we obtain marginally different values for the left and right confidence bounds. In Figure~\ref{fig:accuracy} we show the histograms of the average CI$_{68}$ bounds for both extinction and effective temperature, as well as $\Delta$. Mean values of the confidence intervals are $\left<\text{CI}_{68}(A_0)\right>=\unit{0.04}{mag}$, $\left<\text{CI}_{68}(T_{\rm eff})\right>=\unit{58}{K}$ and $\left<\text{CI}_{68}(\Delta)\right>=\unit{2.0}{mag}$. We can therefore be confident that our parameter estimates are generally precise in a statistical sense.

These values of precision are significantly smaller than the accuracy of the model (Section~\ref{subsec:effectcolours}), which could be taken to mean that the parameter inference is overconfident compared to the information content of the data. The intrinsic scatter of the spline fit is of the same order of magnitude as the photometric errors of the data (see Section~\ref{subsec:buildfm} and Figure~\ref{fig:fmtpsresiduals}), therefore properly representing the variation and uncertainties of the model. 

Average photometric errors in all nine bands vary from $0.01$ to $\unit{0.03}{mag}$, the mean temperature uncertainty in the training data is $\unit{60}{K}$, uncertainties in extinction can only arise from an erroneous correction of the SFD estimates, unless the reddening law is wrong. These, though, are typically only $A_{\rm SFD, r}=\unit{0.03}{mag}$. Our precision therefore is in the order of the parameter errors. If we fix $\sigma_{p_X} = \unit{0.1}{mag}$, which is roughly five times larger than the photometric errors, we obtain precisions of $\left<\text{CI}_{68}(A_0)\right>=\unit{0.36}{mag}$ and $\left<\text{CI}_{68}(T_{\rm eff})\right>=\unit{402}{K}$. These values are in the order of $\unit{25}{\%}$ larger than the MAE reported in Table~\ref{tab:residuals}. The corresponding biases, mean absolute errors and root mean squares only change by a few percent in respect to the tabulated results. Therefore, a more accurate account for model limitations could imply using significantly larger errors on photometry, although on average the statistical performance remains compatible in either case.

\begin{figure}
\includegraphics[width=\columnwidth]{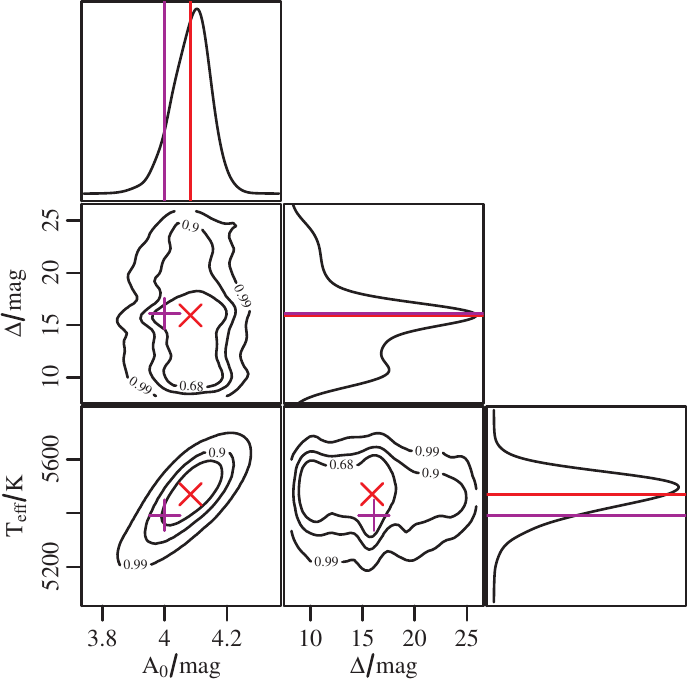}
\caption{Contours of the posterior PDF for a star with true APs of $A_{\rm 0, true}=\unit{4}{mag}$ and $T_{\rm eff, true} = \unit{5392}{K}$. In each panel the red cross or line shows the position of the estimated parameter value ($A_0 = \unit{4.06}{mag}$, $T_{\rm eff} = \unit{5446}{K}$, $\Delta = \unit{15.9}{mag}$), whereas the purple plus or line is the true (constructed) value. Contours show $68\%$, $90\%$ and $99\%$ confidence intervals. The smaller side panels show the marginalised one-dimensional PDFs for the single parameters, as aligned with the respective axes. The peak densities are scaled to $1$ in each case. The indicated true value of distance modulus is estimated using the true temperature and the HRD from section \ref{subsec:hrdprior}. The best estimate values for temperature and extinction are computed using the mean of the distribution, whereas the mode is used for distance modulus.}
\label{fig:contour1}
\end{figure}

To illustrate the output from the Monte Carlo sampling, Figure~\ref{fig:contour1} shows the contours and one-dimensional PDFs of the three parameters extinction, effective temperature and $\Delta$ for a single example star. In the extinction-temperature plane (bottom left panel) we clearly see the degeneracy between the two parameters. In the one-dimensional PDF for $\Delta$ (middle panel) we see the typical bimodal distribution, which is the result of sampling over the parts in the HRD that represent the main sequence and the giant branch. One peak is usually higher, which is why we use the mode to quantify this parameter rather than the mean. In most cases the higher peak relates to the main sequence (though not in this example), which is a reasonable justification to use it, as we have trained our models using photometry from dwarf stars.

It must be noted that the confidence intervals for $\Delta$ are large, though the inner $68\%$ are more tightly constrained. We can reduce this spread by either running a longer Monte-Carlo chain, which adds more weight to the non-background regions in the HRD, or by directly decreasing the background level by a few orders of magnitude when computing it in the first place. Both options result in nearly unchanged parameter estimates but improved precision in $\Delta$. The second option is clearly more appealing, as it does not increase the computation time.

\subsection{Effects of $\bmath{R_0}$}
\label{subsec:r}
The results shown so far were for a fixed $R_0 = 3.1$. In reality, this value is only the mean value for the diffuse interstellar medium, whereas denser regions will typically be characterised by a higher value of $R_0$, arising from a change in size distribution and composition of the dust grains responsible for extinction in optical bands. This does not, however, equate high line-of-sight extinction with large $R_0$; it simply describes properties of the local dust.

We train a new forward model in which we simulate the change of colours due to extinction and $R_0$ for stars with known temperatures. We expand our parameter grid and again use $4\,000$ stars to train and validate the model. The general setup remains the same as before. We use $R_0$ in the range from $2.1$ to $5.9$. The model performance we achieve like this is presented in Table~\ref{tab:residualsr}.

\begin{table}
\caption{Performance using all colours and variable $R_0$, characterised by bias, mean absolute error (MAE) and root mean square (RMS) of the residuals, i.e. the difference between estimated and true parameters, as detailed in section~\ref{subsec:r}.}
\label{tab:residualsr}
\centering
	\begin{tabular}{cccc}
				& $A_0$/mag	& $T_{\rm eff}$/K	& $R_0$	\\ \hline
   		bias		& $-0.22$	& $-206$		& $0.32$ \\
		MAE		& $0.39$		& $438$		& $0.73$ \\
		RMS		& $0.69$		& $771$		& $0.95$
  \end{tabular}
\end{table}

Comparing case (a) in Table~\ref{tab:residuals} with Table~\ref{tab:residualsr}, we see that modelling $R_0$ slightly deteriorates the performance in the other two parameters, the standard deviation and MAE in the residuals are increased by roughly $50\%$ for temperature and extinction. This comes as no surprise, as we are now extracting more parameters, but with the same initial data, noting that the observed change in colours are quite small for larger differences in this parameter. In general we are able to roughly infer $R_0$, though the accuracy is not sufficient to confidently differentiate between, say, $3.1$ and $3.8$. We therefore build the lookup table using somewhat larger steps of $0.2$ in $R_0$ for the range noted above, and the other parameters remaining the same.

\begin{figure}
\includegraphics[width=\columnwidth]{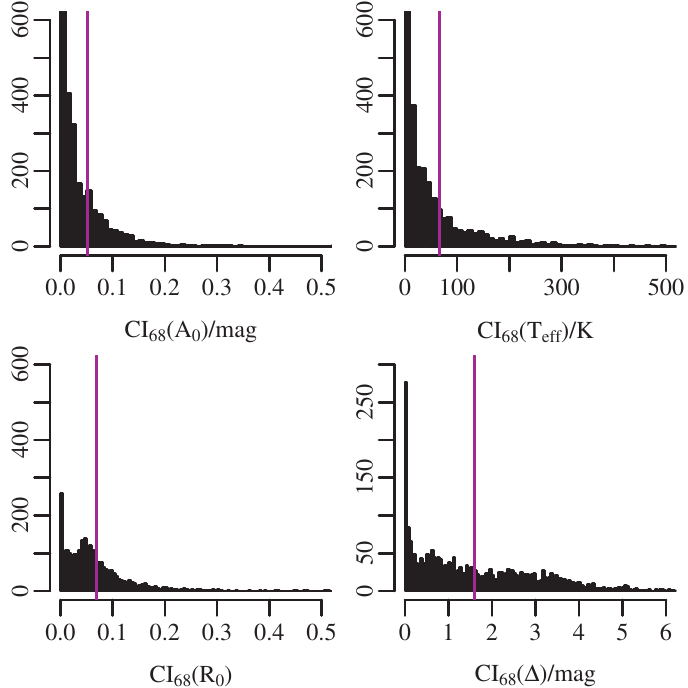}
\caption{Histogram of $68\%$ confidence intervals of extinction, temperature and extinction factor, which are estimates of the uncertainty on each parameter we infer. The purple vertical lines indicate the mean values in each case. For $A_0$ this corresponds to $\unit{0.05}{mag}$, for $T_{\rm eff}$ it is $\unit{66}{K}$, for $R_{0}$ it is $0.07$ and for $\Delta$ it is $\unit{1.6}{mag}$.}
\label{fig:accuracyr}
\end{figure}

In Figure~\ref{fig:accuracyr} we present the precision of the predicted parameters as histograms, where the purple lines again depict the mean values. We obtain $\left<\text{CI}_{68}(A_0)\right>=\unit{0.05}{mag}$, $\left<\text{CI}_{68}(T_{\rm eff})\right>=\unit{66}{K}$, $\left<\text{CI}_{68}(R_0)\right>=0.07$ and $\left<\text{CI}_{68}(\Delta)\right>=\unit{1.6}{mag}$. The first two values are minimally larger than in the previous case, though it is to be expected considering that we have added an additional degree of variation to the model. Again, the confidence intervals are significantly smaller than the model accuracy.

\section{Application to multiple fields}
\label{sec:application}
We have so far summarised the accuracy and precision of the model using data with known stellar parameters. We now apply the model to estimate APs for large areas of the sky where individual estimates are not available. Querying the \textit{WFCAM Science Archive} with the flags detailed in section \ref{subsec:training}, but imposing no constraints on SSPP information, we obtain a total of $4\,191\,659$ unique stellar objects, of which $3\,055\,954$ and $1\,135\,705$ are north and south of the Galactic equator, respectively. We also pick out a section of the southern sky with $l > 180^\circ$, which we use to illustrate the variations of the model, namely using only photometry from SDSS, using all photometry, and also including $R_0$ in the parameter estimation.

\begin{figure*}
\includegraphics[width=\textwidth]{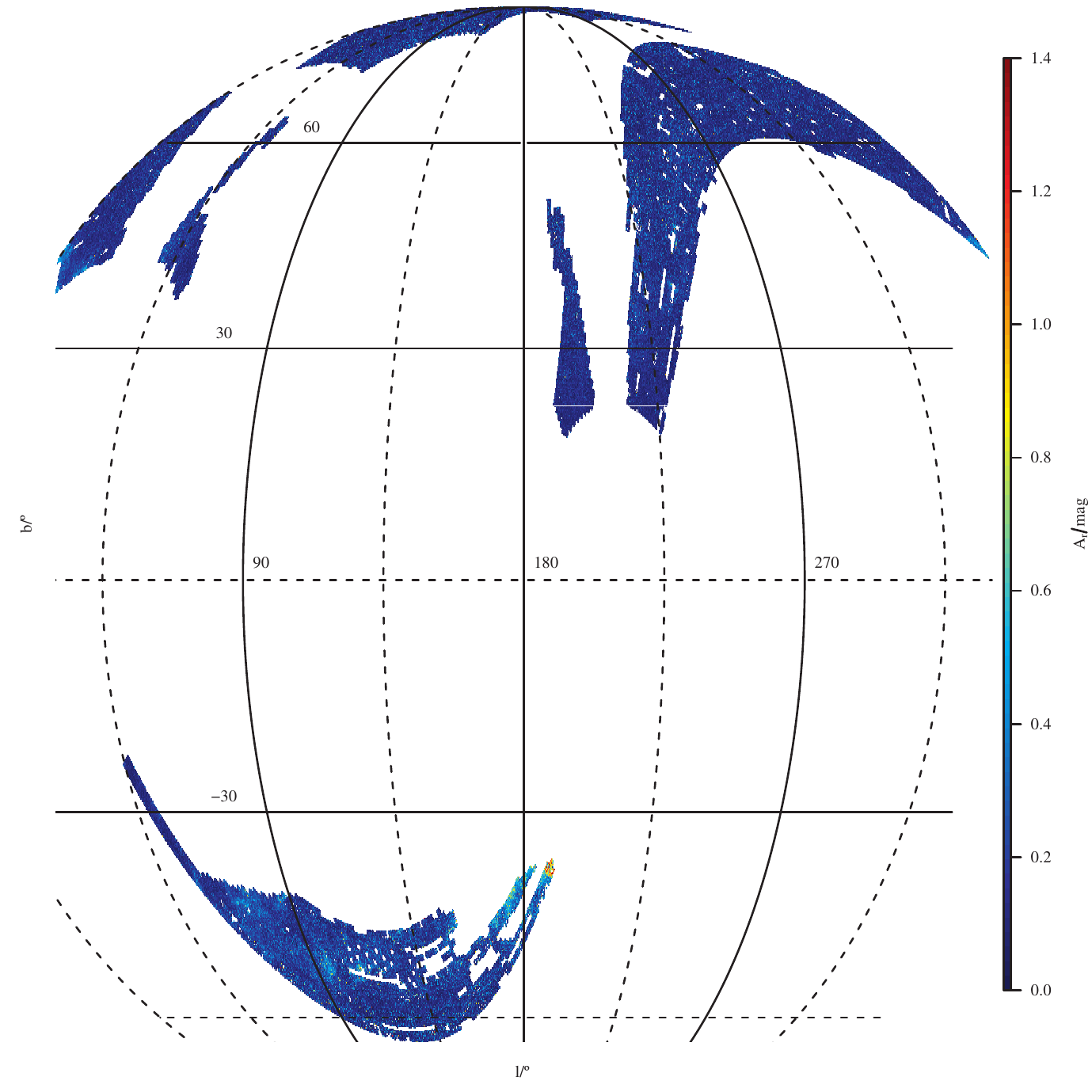}
\caption{A Mollweide equal area projection of the mean extinction in the full crossmatched portion of the sky, centred on the Galactic anti-centre. The stars are binned using a resolution of $\unit{11.48}{arcmin/pixel}$ in $l$ and $b$. The non-cartesian grid in latitude and longitude is overplotted. The colour scale shows the mean extinction $A_{\rm r}$ in any given pixel. White areas are not jointly covered in the surveys.}
\label{fig:fullsky}
\end{figure*}

We first show the full projected maps of the northern and southern part of the surveys in Figures~\ref{fig:fullsky}. Here we use a Mollweide projection in Galactic longitude and latitude to obtain equal area pixels, using a resolution of $\unit{11.48}{arcmin/pixel}$ in both coordinates. This corresponds to an average of $29$ stars per pixel, depending on the region observed, with extremes of $3$ and $250$ stars per pixel.

The interstellar medium towards the Galactic poles typically hosts diffuse gas and lines of sight with low extinction. Nevertheless, the area jointly probed by the two surveys does cover some regions with higher average extinction. Generally we can pick out larger structures that are also visible in SFD maps of the same resolution. Towards the right of the southern part of the sky one can start to make out parts of a larger structure, with extinction estimates in the range of $1$ to $\unit{1.5}{mag}$ in the $r$-band.
 
\begin{figure}
\includegraphics[width=\columnwidth]{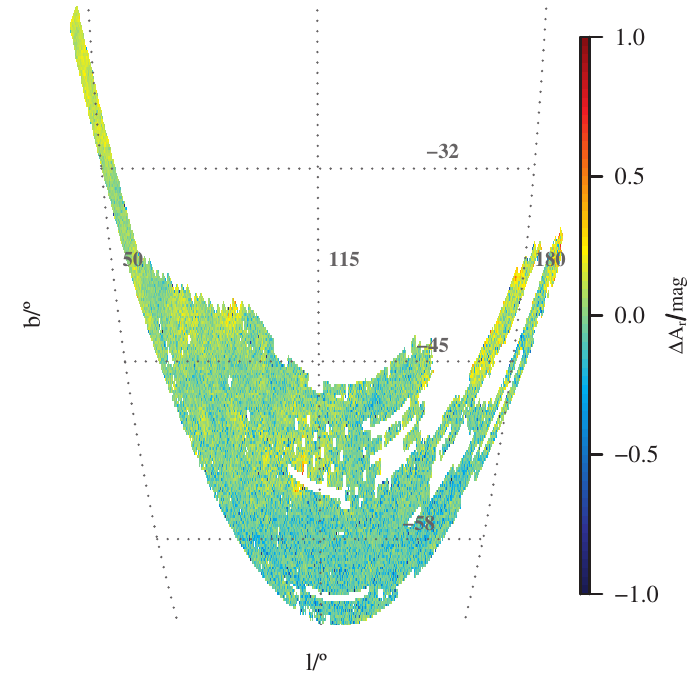}
\caption{Difference between SFD extinction estimates and our own for stars in the southern part of the sky, using a Mollweide projection with $\unit{11.48}{arcmin/pixel}$ resolution. The colour coding in the range of $\Delta A_{\rm r} = -1$ to $\unit{1}{mag}$ shows regions where SFD predicts a higher extinction in yellow/red (positive), and where our model has a higher estimate in blue (negative).}
\label{fig:southsfddiff}
\end{figure}

\begin{figure}
\includegraphics[width=\columnwidth]{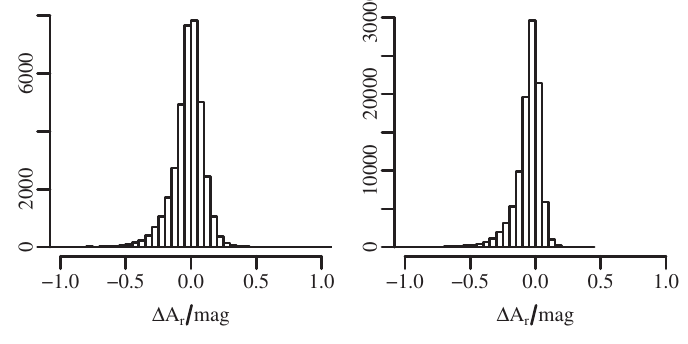}
\caption{Histograms of differences between SFD extinction estimates and our own for stars ($\Delta A_{\rm r} = A_{\rm r, SFD} - A_{\rm r}$) in the southern part of the sky (left) and in the northern (right).}
\label{fig:sfddiffhist}
\end{figure}

To more quantitatively assess the credibility of the presented map, we compare the computed extinction values with those from SFD. For this we project the data in an identical manner and calculate the difference between the SFD extinction values and our own estimates. This is plotted in Figures \ref{fig:southsfddiff} for the southern part of the sky. We refrain from showing the equivalent image for the northern part, as it qualitively shows the same behaviour. In Figure~\ref{fig:sfddiffhist} we show the histograms for the differences between SFD extinction estimates and our own for both the southern and northern part of the sky, based on the binned data. In the south the mean difference and standard deviation are $\unit{-0.02}{mag}$ and $\unit{0.13}{mag}$, respectively. In the north we obtain $\unit{-0.06}{mag}$ and $\unit{0.11}{mag}$.

The general trend favours very small differences, with the mean $\Delta A_{\rm r}$ being close to zero. Some regions indicate that SFD predicts higher extinction than we do. This can have several reasons, one being the fact that we compute extinction along a line of sight to individual stars, rather all the way to the edge of the Galaxy. Then again, we are looking at high (absolute) latitudes, where we assume extinction in general to be small, allowing us to cover most of the stars. Selection effects and limitations set by our data retrieval could affect this assumption. The few pixels that have a blue hue, indicating that our model predicts higher extinction, are mostly artificial effects that arise from binning the data, leaving only very few stars in those bins. If these stars happen to have (erroneously) high extinction estimates, they may not be averaged out.

\begin{figure}
\includegraphics[width=\columnwidth]{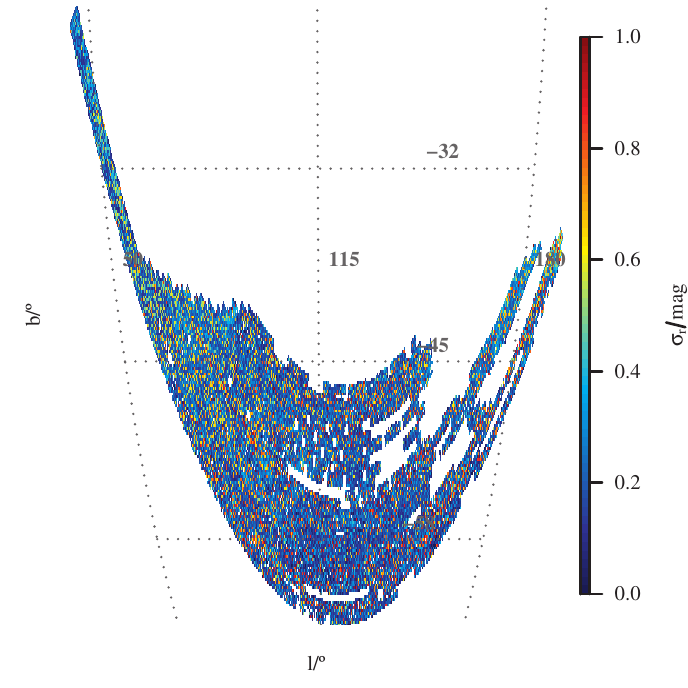}
\caption{Standard deviation $\sigma_{\rm r}$ of $r$-band extinction over the southern sky using nine photometric bands and fixed $R_0$. The colour axis is constrained from $0$ to $1$, only $\unit{0.5}{\%}$ of the cells have higher values.}
\label{fig:southsd}
\end{figure}

\begin{figure}
\includegraphics[width=\columnwidth]{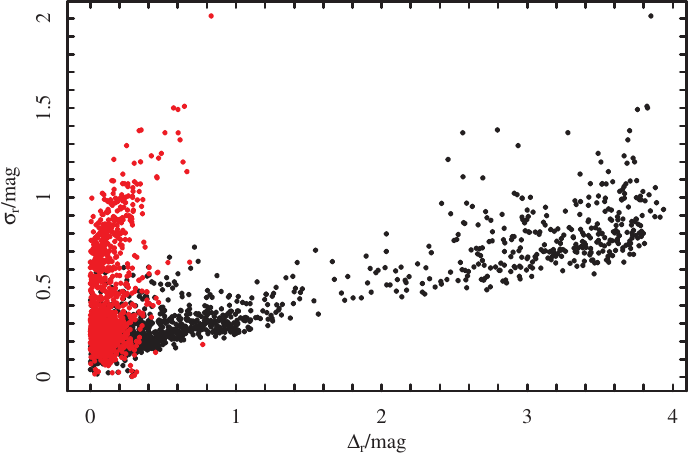}
\caption{Standard deviation $\sigma_{\rm r}$ of $r$-band extinction as a function of $\Delta_{\rm r} = |A_{\rm r} - A_{\rm r, SFD}|$, the difference between our estimates and SFD values for each pixel in the southern sky using nine photometric bands and fixed $R_0$. Red dots are used for mean extinction estimates, whereas black dots use the maximum extinction value in each pixel (see Figure~\ref{fig:southl180upsmax}).}
\label{fig:southsd_vs_resid}
\end{figure}

This is visualised in Figure~\ref{fig:southsd}, where we show the standard deviation of $A_{\rm r}$ for each pixel over the southern part of the sky. We see a large range of estimates in each pixel, independent of the actual mean extinction in them and the position on the sky. There is a noticeable correlation between the standard deviation and the (absolute) difference between our extinction estimate and that of SFD, $\Delta_{\rm r} = |A_{\rm r} - A_{\rm  r, SFD}|$, see Figure~\ref{fig:southsd_vs_resid}. Red dots are computed using the mean extinction estimates for $A_{\rm r}$ in the equation above. Black dots use the maximum extinction value in each pixel, i.e. $A_{\rm r, max}$ (see Section~\ref{subsec:furtherpriors} and Figure~\ref{fig:southl180upsmax} for further details). In this case the range of residuals is naturally larger. In both cases, though, the standard deviation in a pixel is a function of the difference between the two extinction estimates.

On the whole, the results are consistent with how we expect the model to function, with results obtained from SFD or other data that deliver the integrated extinction to the edge of the Galaxy. On average, we slightly underestimate the mean extinction in a pixel due to the fact that we are averaging over multiple stars with varying degrees of precision in their individual estimates.

\subsection{Extinction prior}
\label{subsec:furtherpriors}
A simple prior to implement is using SFD estimates to constrain maximum extinction along any line of sight. Though the SFD maps do have several systematic problems, overestimating extinction particularly towards the Galactic plane and in regions of high extinction \citep[e.g.][]{2010ApJ...725.1175S, 2011ApJ...737..103S}, at the latitudes probed here the corrections are only on the percent level. By constructing a broader prior we can safely ignore this.

In practice we implement the prior as a step function with the probability dropping from $1$ to $0$ at $1.3\cdot A_{\rm r, SFD}$ with $A_{\rm r, SFD}$ being the converted SFD $r$-band extinction. This factor of $1.3$ allows for enough range in the extinction estimates to account for errors in the reference value and smoothing of the data, whilst also not being too restrictive. To compare the performance of a model using this prior we use the known, simulated extinction. We see only negligible changes to the values in Table~\ref{tab:residuals} when adding the prior. Naturally, more stars remain within the training grid, as maximal values in extinction are constrained. Considering average parameter estimates do not change much, we can be confident that the model can properly characterise the degeneracy intrinsic to the data, independent of stronger priors.

\begin{figure*}
\includegraphics[width=\textwidth]{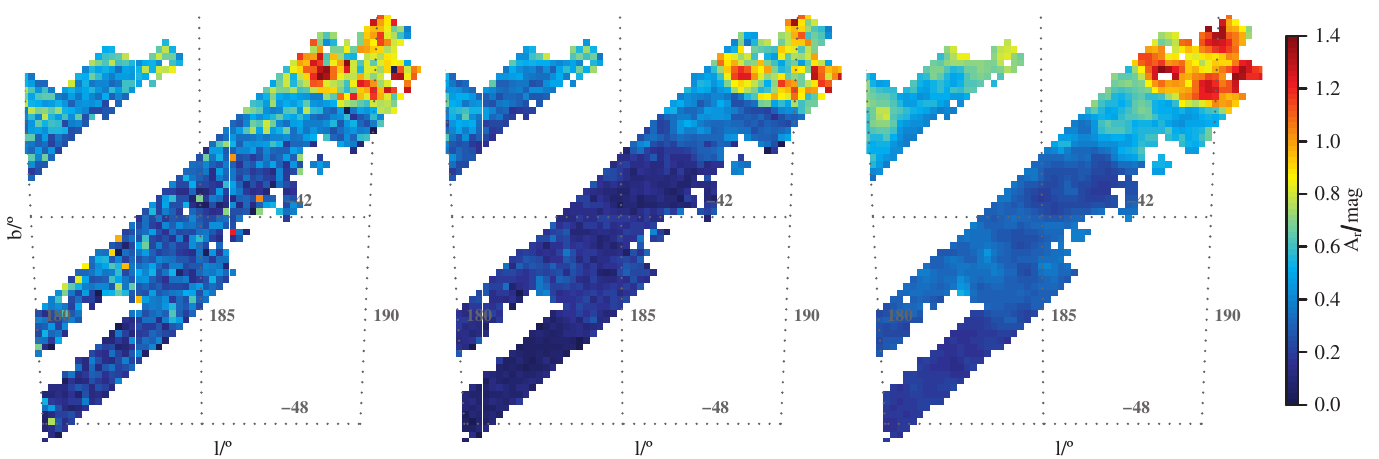}
\caption{Extinction maps for a subset of the data in the southern sky in a Mollweide projection. The non-cartesian grid in latitude and longitude is overplotted. The data are binned to a resolution of $\unit{11.48}{arcmin/pixel}$. The left panel shows the mean $r$-band extinction in each pixel for the case of using no extinction prior, the middle panel that for a prior based on SFD estimates. In both cases eight colours are used with fixed $R_0$. The right panel shows for comparison the same region using the SFD estimates for the direction of each pixel directly. White areas are either not covered by the data or are missed due to binning (due to post-processing, the panels do not necessarily use an identical set of stars, therefore some pixels appear white in the left panel but not in the right two.)}
\label{fig:southl180ups}
\end{figure*}

When applying this updated model to data (i.e. with unknown APs), we do notice some changes in respect to average extinction estimates. In Figure~\ref{fig:southl180ups} we show the differences for a region of the southern sky. The data are projected using a Mollweide projection and colour-coded according to the mean extinction in each pixel. The left panel depicts the results without priors (fixed $R_0$ and eight colours). In the right panel, the equivalent map of SFD estimates is shown, and the centre panel that of our model including the SFD-based prior on extinction. Clearly, inclusion of the prior reconstructs more closely the SFD reference map, whereas the alternative produces a less smooth map, although it still closely follows the depicted distribution on a whole, picking out the high extinction regions towards the top-right, as well as the general trends. Individual pixels are not always accurate though. This is due to the fact that we estimate the APs for every star individually, whereas the smoothed SFD map is a proxy for the average extinction along a line of sight.

\begin{figure}
\includegraphics[width=\columnwidth]{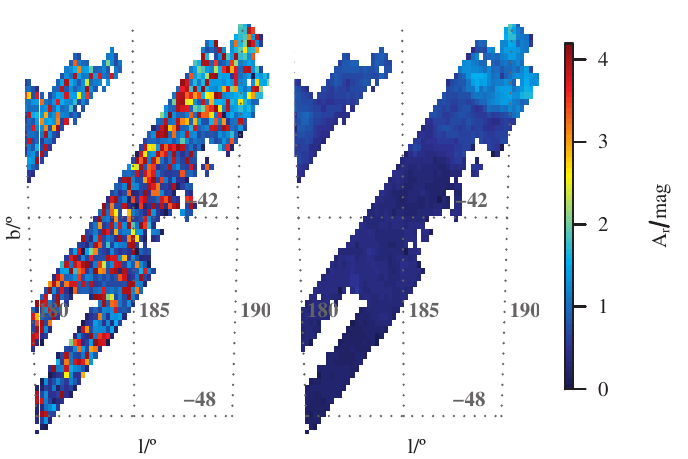}
\caption{Region and resolution as in Figure~\ref{fig:southl180ups}. The left panel shows the maximum $r$-band extinction in each pixel for the case of using no extinction prior, the right panel that for a prior based on SFD estimates. In both cases eight colours are used with fixed $R_0$. Note that the colour scale has changed to cover the parameter range.}
\label{fig:southl180upsmax}
\end{figure}

This issue becomes more pronounced when we use the maximal value in any bin to create the map. This is shown in Figure~\ref{fig:southl180upsmax}, where we project the maximum values for the case of using the standard model (left) and including the prior (right). The equivalent map for SFD values is identical to the right panel in Figure~\ref{fig:southl180ups} as the reported values are smoothed already. With this prior we limit the maximum value for each line of sight. The resulting map is almost identical to the SFD reference map, with slight differences visible particularly at the edges of the footprints due to strongly varying stellar number densities. Ignoring the prior we compute varying maximum values of extinctions, as we have no low boundary limit. As can be seen in Figure~\ref{fig:southl180ups}, when using the mean these extreme values tend to average out, even though individual estimates may still miss the true (or expected) value.
 
\subsection{Extinction profiles}
\label{subsec:distanceprofiles}
In addition to extinction and temperature we have inferred distances to all stars. We also have full probabilistic information available for each parameter which we can use to compute profiles of extinction as a function of distance for individual lines of sight.  The weighted mean $\left< A_{\rm r} \right>$ and standard deviation $\sigma_{A_{\rm r}}$ for each distance modulus point $\mu$ are computed as
\begin{equation}
\left< A_{\rm r}(\mu) \right> = \frac{\sum A_{{\rm r}, i} w_i}{\sum w_i} \quad , \quad
\sigma_{A_{\rm r}} = \sqrt{\frac{\sum w_i (A_{{\rm r}, i}-\left< A_{\rm r} \right>)^2}{\frac{N-1}{N}\sum w_i}}.
\label{eq:weightedmean}
\end{equation}
All stars whose $68\%$ confidence intervals extends to the distance modulus selected are included. The standard deviation of each individual measurement is an order of magnitude smaller and barely affects the size of the total standard deviation compared to that of the generic weighted mean stated above. The weight $w_i$ for each star is computed using the difference between $\mu_i$ and $\mu$. We compute the probability for the normalised, asymmetric Gaussian described by the mode and left and right $68\%$ confidence intervals at that difference. With $U = \mu_i - \mu$ we obtain
\begin{equation}
w_i = \frac{2}{\sqrt{2\pi}(\sigma_1+1)}\cdot\exp\left( -\frac{U^2}{2\sigma_1^2}\right) \ ,
\label{eq:weights}
\end{equation}
where $\sigma_1$ is the smaller of the two confidence intervals. Stars that have a more precise estimate of distance modulus (small $\sigma_1$) are weighted more strongly than those with imprecise estimates (large $\sigma_1$). We repeat this for every (arbitrary) value of distance modulus we would like to use. This way we can compute extinction profiles using the full PDF of each star instead of merely averaging the data directly. Similarly this can be done for $A_0$ or proper distance too.

\begin{figure}
\includegraphics[width=\columnwidth]{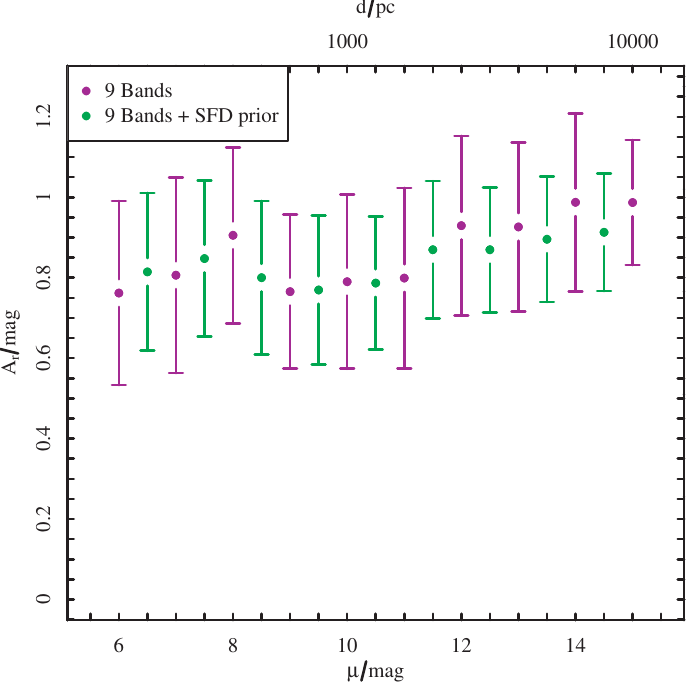}
\caption{Mean $r$-band extinction as function of distance modulus for $520$ stars in the region $l \in \unit{(189.56, 190.88)}{^\circ}$ and $b \in \unit{(-37.35, -36.35)}{^\circ}$. Mean values and error bars are computed as per Equations~\ref{eq:weightedmean} and \ref{eq:weights}. The average SFD extinction for this field is $A_{\rm r, SFD} = \unit{1.16}{mag}$. Both profiles use all nine bands, purple points show the case of no prior and green points including a prior on extinction (Section~\ref{subsec:furtherpriors}). Top axis shows the corresponding distances $d = 10^{\mu/5+1}$ in parsec.}
\label{fig:distprofile}
\end{figure}

An example is shown in Figure~\ref{fig:distprofile} for the region $l \in \unit{(189.56, 190.88)}{^\circ}$ and $b \in \unit{(-37.35, -36.35)}{^\circ}$ (top right section of Figure~\ref{fig:southl180ups}). We have chosen this region because it has higher extinction and as such can demonstrate the application of Equations~\ref{eq:weightedmean} and \ref{eq:weights}. The figure shows two profiles, both using the nine band forward model, one without prior (purple) and one including a prior on extinction as in Section~\ref{subsec:furtherpriors} (green). The profiles are similar in the sense that they both are essentially flat, possibly with a slight increase towards further distances, though the error bars on extinction are compatible with both interpretations. The errors on distance modulus are implicitly included in the way the mean extinction and its error are computed (see Equations~\ref{eq:weightedmean} and \ref{eq:weights}). The average precision of distance modulus estimates varies from approximately $\unit{0.5}{mag}$ to $\unit{1.5}{mag}$ when going from the smallest to the largest value of $\mu$ in Figure~\ref{fig:distprofile}. The SFD extinction for this field is $A_{\rm r, SFD} = \unit{1.16}{mag}$, which is covered by the asymptotic error bars. The green curve tends to be slightly smaller, in line with conclusions from the previous section, where the maximal extinction estimates are constrained by the prior. However, the values are compatible within their uncertainties. Despite imposing no constraints on distance modulus when inferring the APs (apart from fitting into the parameter space of the HRD) and when computing the profiles, the variation of extinction with distance is physically consistent with extinction not decreasing with increasing distance. Furthermore, the shapes of the profiles are in agreement with the assumption that the stars at high latitudes reside behind the dust layers. Given the bright magnitude limit of the surveys used, we do not expect to find stars with distance moduli less than $\mu = \unit{6}{mag}$. Indeed, our method estimates very few stars to have smaller values of $\mu$.

Lines of sight in regions with lower extinction behave in a similar manner, though naturally with lower asymptotic values of the extinction. In particular, cells at small distances show larger scatter, which can be attributed to low sampling statistics at these values, with only a few stars being computed to be very close by. In contrast, at larger distances the extinction in each cell is computed using hundreds of stars (owing to large uncertainties in the distance estimation and roughly $d^3$ growth in the volume of a cell with distance).
\subsection{Inferring $R_0$}
\label{subsec:r0}
\begin{figure}
\includegraphics[width=\columnwidth]{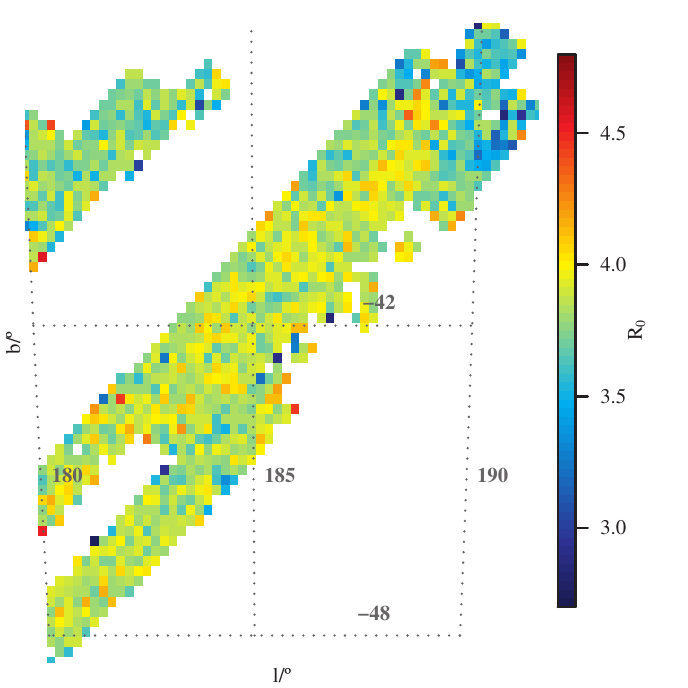}
\caption{Distribution of the parameter $R_0$ over a subsection of the southern sky with an average over the whole field of $\left<  R_0 \right> = 3.75\pm0.19$. The same resolution as in previous maps is used.}
\label{fig:l180r0}
\end{figure}

As detailed in Section~\ref{subsec:r} we are able to infer the $R_0$ parameter with moderate accuracy, in addition to extinction, temperature and $\Delta$. Using the corresponding forward model, we reran the method on the same small field of stars as before. We refrain from showing the projected extinction map as it does not differ qualitatively from that with constant $R_0$ (Figure~\ref{fig:southl180ups} left and centre panels). We instead show in Figure~\ref{fig:l180r0} the mean distribution of $R_0$ in the same projection.

Generally we see quite some scatter around the mean $\left<  R_0 \right> = 3.81\pm0.20$ in this field. Using the raw data (i.e. not binned) we obtain $\left<  R_0 \right>_{\rm raw} = 3.82\pm0.47$, which has a slightly larger standard deviation about the mean. Noting that we average over a large region with varying degrees of extinction and that the model accuracy is only $\text{MAE}(R_0)=0.74$, this should not be taken as a strong statement concerning the average $R_0$ of the diffuse ISM.

Nonetheless, this supports results by \cite{2012AstL...38...12G,2012AstL...38...87G}, who finds $R_0$ to be quite large and to have significant variations at high latitudes and low extinctions. In comparison, \cite{2012ApJ...757..166B} find $R_0 = 3.0$ with $0.1$ random and systematic uncertainties, although they argue that for $A_{\rm r} < \unit{2}{mag}$, estimates on $R_0$ are very unreliable.

\begin{figure}
\includegraphics[width=\columnwidth]{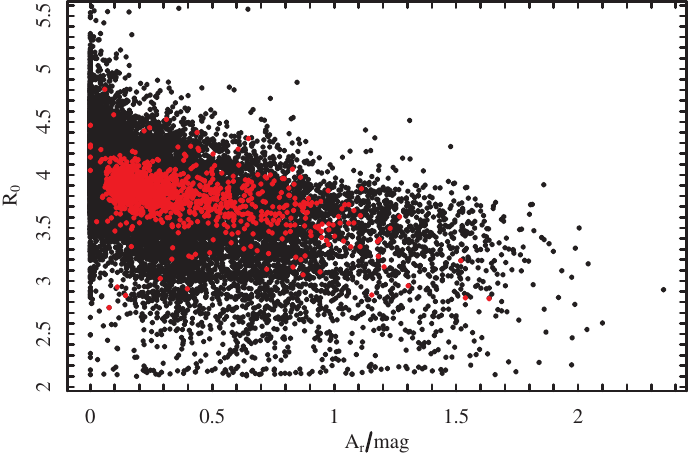}
\caption{For each pixel of the map in Figure~\ref{fig:l180r0} we plot $R_0$ vs. $A_{\rm r}$. Black points are the raw, unbinned data, red are for the binned cells. We see the scatter about the mean $\left<  R_0 \right> = 3.75\pm0.19$ (red) and $\left<  R_0 \right> = 3.75\pm0.47$ (black).}
\label{fig:l180r0vsa0}
\end{figure}

Looking at the relation between $A_{\rm r}$ and $R_0$ (for raw and binned data) we see in Figure~\ref{fig:l180r0vsa0} that there is a trend ($R_0 = (-0.45\pm0.2)A_{\rm r}\text{[mag]} + 4.0$), again consistent with the previous statement in the limit of low extinctions.

\begin{figure}
\includegraphics[width=\columnwidth]{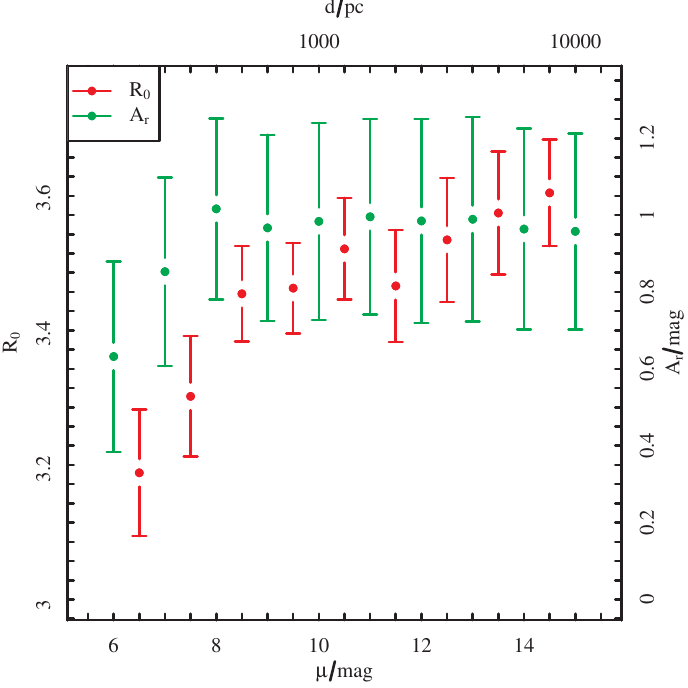}
\caption{Mean $R_0$ (red, left axis) as function of distance modulus for a subset of stars from Figure~\ref{fig:l180r0}. The same technique as in Section~\ref{subsec:distanceprofiles} is used to compute the mean and standard deviation. In green the extinction profile is plotted (right axis). Distances in parsec are given in the top axis.}
\label{fig:l180distprofiler0}
\end{figure}

Using the same technique as in Section~\ref{subsec:distanceprofiles} we compute the profiles of $R_0$ as a function of distance for a small region at the top right of Figure~\ref{fig:l180r0vsa0} (as in Figure~\ref{fig:distprofile}). This is shown in Figure~\ref{fig:l180distprofiler0} in red (left axis), where the mean extinction has been replaced with mean $R_0$ in the equations. In green the extinction profile for the same data is plotted (right axis). As in the fixed $R_0$ case, we see a slight increase in extinction towards larger distances, though it is more pronounced now. At the same time $R_0$ increases quite quickly, although the amplitude of variations are roughly the size of the error bars and well within the model accuracy. To be able to further analyse this interesting aspect, we need to probe regions with higher values of extinction (not covered by the crossmatched SDSS/UKIDSS data) or obtain more informative data from other sources. As it stands, we can only make the general statements mentioned above.

\section{Conclusions}
\label{sec:summary}
Combining large optical and near infrared surveys of stars, such as SDSS and UKIDSS, is an effective means of constraining interstellar extinction as well as addressing the degeneracy between effective temperature $T_{\rm eff}$ and extinction $A_0$ for individual stars. To achieve this, we construct a forward model trained on colour changes of real data due to extinction, effective temperature and $R_0$ in combination with self-consistent use of an HRD prior to infer distance information. Incorporating these physical constraints, we infer astrophysical parameters (APs) in a Bayesian framework using an MCMC algorithm to efficiently sample over the posterior distribution. This way we are able to naturally extract full probability distribution (PDF) information in each AP. Testing with synthetic data has shown that this method produces accurate results, with the caveat of not fully incorporating the systematic effect metallicity has on distance estimation.

We have investigated the impact of the data and assumptions on the model, in particular the effect of removing photometric bands, inferring $R_0$ in addition, and the effect of an incorrectly assumed metallicity on the results. When using nine bands (five from SDSS, four from UKIDSS) we obtain a mean absolute error of the residuals of the model for $A_0$ of $\unit{0.23}{mag}$ and $\unit{300}{K}$ in $T_{\rm eff}$. These values are up to $50\%$ better than just using SDSS photometry. The MAE for $\Delta = m_{\rm r} - M_{\rm r}$ is $\unit{2.2}{mag}$. With the same set of data we can also achieve an accuracy of $0.73$ in $R_0$, with acceptable changes to the performance in other parameters. Using the $68\%$ confidence intervals (CI) to quantify the individual precision of the AP estimates, we obtain the following values: $\left<\text{CI}_{68}(\Delta)\right> = \unit{1.6}{mag}$, $\left<\text{CI}_{68}(A_0)\right> = \unit{0.05}{mag}$, $\left<\text{CI}_{68}(T_{\rm eff})\right> = \unit{66}{K}$, $\left<\text{CI}_{68}(R_0)\right> = 0.07$. When estimating APs for fixed $R_0$ the precision improves slightly. Accuracies of distance modulus (or $\Delta$) estimates are strongly dependent on the correct assignment of metallicity to the stellar population being analysed. This can either be achieved by matching the metallicity of the HRD prior, or by directly inferring metallicity along with the other parameters. This, however, requires higher quality and more informative data or spectroscopic observations of the stars.

Although we estimate APs for stars individually, our method is able to trace the intermediate-size dust structures visible in e.g. SFD dust maps of the same regions. In addition to line of sight extinction estimates (either averaged or per star) we can use the full PDFs to compute probabilistic profiles of $A_0$ and $R_0$ as a function of distance, indicating how dust is distributed along the line of sight. We generally find the expected result that, at these Galactic latitudes, the observed stars are behind the layers of dust.

We have limited this work to stars at high Galactic latitudes common to SDSS and UKIDSS Large Area Survey, allowing us to obtain photometry in nine bands. In principle this method can use fewer bands, with the decrease in accuracy detailed above, or use data from other surveys, such as Pan-STARRS \citep{2002SPIE.4836..154K} or 2MASS \citep{2006AJ....131.1163S}. This would allow us to probe high extinction regions towards the Galactic plane and further address the performance in respect to $R_0$.

As analysed in \cite{2011MNRAS.411..435B}, the method can be expanded to combine distance estimates, such as parallaxes, with the HRD prior in order to provide a more accurate estimate of APs. Furthermore, metallicity and surface gravity may be estimated as well, although, as noted above, this requires higher quality data. But if we could estimate them using spectroscopy or more photometric bands, then this would alleviate the strong assumption made by implementing an HRD with fixed metallicity, and thus potentially improve the accuracy of distance estimates.

Current work for the Gaia \citep{2001A&A...369..339P} data processing pipeline, as summarised in \citet{2013arXiv1309.2157B}, illustrate AP estimation performance using Gaia spectra and photometry. By combining current multiband surveys with Gaia parallaxes we expect to be able to increase the accuracy of the method and precision of the parameter estimates significantly.

\section*{Acknowledgments}
We thank E.~F. Schlafly and R. Shetty for helpful comments and discussions. We also thank an anonymous referee for detailed comments on the paper. We would also like to thank an anonymous referee for helpful comments.

This project is funded by the Sonderforschungsbereich SFB881 'The Milky Way System' (subproject B5) of the German Research Foundation (DFG). RJH is member of the International Max-Planck Research School for Astronomy and Cosmic Physics at the University of Heidelberg (IMPRS-HD) and the Heidelberg Graduate School of Fundamental Physics (HGSFP). 

This work makes use of the ninth data release of UKIDSS. The UKIDSS project is defined in \citet{2007MNRAS.379.1599L}. UKIDSS uses the UKIRT Wide Field Camera \citep[WFCAM;][]{2007A&A...467..777C}.

Funding for SDSS-III has been provided by the Alfred P. Sloan Foundation, the Participating Institutions, the National Science Foundation, and the U.S. Department of Energy Office of Science. The SDSS-III web site is http://www.sdss3.org/.

SDSS-III is managed by the Astrophysical Research Consortium for the Participating Institutions of the SDSS-III Collaboration including the University of Arizona, the Brazilian Participation Group, Brookhaven National Laboratory, University of Cambridge, Carnegie Mellon University, University of Florida, the French Participation Group, the German Participation Group, Harvard University, the Instituto de Astrofisica de Canarias, the Michigan State/Notre Dame/JINA Participation Group, Johns Hopkins University, Lawrence Berkeley National Laboratory, Max Planck Institute for Astrophysics, Max Planck Institute for Extraterrestrial Physics, New Mexico State University, New York University, Ohio State University, Pennsylvania State University, University of Portsmouth, Princeton University, the Spanish Participation Group, University of Tokyo, University of Utah, Vanderbilt University, University of Virginia, University of Washington, and Yale University.

\label{lastpage}

\end{document}